\documentstyle[aipmod,eqalign,graphicx,times]{article}
\advance \textwidth by 1in
\advance \oddsidemargin by -0.5in
\advance \textheight by 1in
\advance \topmargin by -0.5in
\let\sc=\small
\font\tenmib=cmmib10 scaled \magstep1 \font\tensfb=cmssbx10 scaled \magstep1 
\def\dv{d^3{\bf v}}\def\da{d^2\!{\bf A}}\def\dvp{d^3{\bf p}}
\def\bphi{\hbox{\tenmib \char'036 }}
\def\bsigma{\hbox{\tenmib \char'033 }}
\def\abs#1{\left|#1\right|}

\def\mat#1{\hbox{\tensfb #1}}
\def\secref#1{sec.~\ref{#1}}
\def\secsref#1{secs.~\ref{#1}}\def\Secsref#1{Sections~\ref{#1}}
\def\appref#1{appendix~\ref{#1}}

\def\eqref#1{eq.~(\ref{#1})}\def\Eqref#1{Equation~(\ref{#1})}
\def\eqsref#1{eqs.~(\ref{#1})}
\def\figref#1{fig.~\ref{#1}}
\def\tabref#1{table~\ref{#1}}
\def\Eqlab#1{\eqn(\eqlab{#1})}
\def\refref#1{ref.~\citenum{#1}}
\def\grapprox{\mathrel{\rlap{\lower3.5pt\hbox{$\mathchar"218$}}\raise 1pt
				\hbox{$\mathchar"13E$}}}
\def\lsapprox{\mathrel{\rlap{\lower3.5pt\hbox{$\mathchar"218$}}\raise 1pt
				\hbox{$\mathchar"13C$}}}
\def\grls{\mathrel{\rlap{\lower2.5pt\hbox{$\mathchar"13C$}}\raise2.5pt
						\hbox{$\mathchar"13E$}}}
\def\lsgr{\mathrel{\rlap{\lower2.5pt\hbox{$\mathchar"13E$}}\raise2.5pt
						\hbox{$\mathchar"13C$}}}
\def\fract#1/#2{{\textstyle{#1\over #2}}}
\def\half{\fract1/2}
\begin{document}
\date{PPPL--2290 (Nov.~1985)\\
Comp.\ Phys.\ Rep.\ {\bf 4}(3--4), 183--244 (Aug.~1986)}
\title{Fokker--Planck and Quasilinear Codes\thanks
{ Presented at the 3rd European Workshop on Problems in the Numerical Modeling
  of Plasmas,  Varenna, Italy, September 10--13, 1985.
  Reprinted in {\em Problems in the Numerical Modeling of Plasmas},
  edited by K. Appert (1986).
}} 
\author{Charles F. F. Karney\\
Plasma Physics Laboratory, Princeton University\\
P.O.\ Box 451\\
Princeton, New Jersey 08544--0451, U.S.A.}
\maketitle

\begin{abstract}
The interaction of radio-frequency waves with a plasma is described by
a Fokker--Planck equation with an added quasilinear term.  Methods for
solving this equation on a computer are discussed.
\end{abstract}
\section{Introduction}\label{intro}

In this paper, I will concentrate on those Fokker--Planck models which
are most useful for the study of rf-driven currents \cite{Fisch1}.  I
will therefore take the plasma to be azimuthally symmetric about the
magnetic field and homogeneous (representative of the central portion
of a tokamak plasma).  The Fokker--Planck equation then reduces to an
equation in time and two velocity (or momentum) dimensions only.  This
simplified model yields a wealth of interesting physics and furthermore
illustrates the main numerical problems encountered in more complicated
situations.  In addition to the collision term, the equation will
include the effects of externally injected rf power via a quasilinear
diffusion term, and a dc electric field.  (The electric field arises
whenever the current is time-varying, e.g., during current ramp-up.)
Because the wave interacts with very fast electrons, relativistic
effects are also considered.  In addition, the adjoint method for
solving for moments of the Fokker--Planck equation is discussed.  This
method allows for a great reduction (by orders of magnitude) in the
amount of computer time required.

The paper is divided into three parts:
In the first part of the paper, I give the formulation of the
Fokker--Planck equation.  In \secref{prelim}, the
Fokker--Planck equation and the coordinate systems are introduced.
The collision operator and approximations to it are given in
\secsref{coll} and \ref{coll-approx}.  Corresponding expressions for 
the  quasilinear diffusion
operator are given in \secref{ql}.  The next part of the paper
describes the numerical solution of the equation.  Its boundary
conditions are considered in \secref{bcs}.  \Secsref{spatial}
and \ref{time} describe the spatial and temporal differencing  of the
equation.  In \secref{steady-state}, we describe techniques for
obtaining the time asymptotic solution to the equation.  The last part
of the paper describes the incorporation of relativistic effects
(\secref{relativity}) and the adjoint method for solving the
Fokker--Planck equation (\secref{adjoint}).

The numerical methods presented here are those used in the
Fokker--Planck code used by the author.  A word about the lineage of
this code is in order:  Fokker--Planck codes were developed at Livermore
by Killeen {\sl et al.}\ \cite{Killeen1,Killeen2} for the study of
mirror-machine plasmas.  The latest stage in the development of these
codes is {\sc FPPAC} \cite{McCoy} which is a two-dimensional
multispecies nonlinear Fokker--Planck package.  The Livermore code was
extensively modified by Winsor and Fallon for the study of runaway
electrons, which was undertaken by Kulsrud {\sl et al.}\ \cite{Kulsrud}.
This code has been used by the author in various studies of current
drive beginning with lower hybrid current drive \cite{Karney-lh}.  Over
the years several further modifications have been made, although the
basic structure of the code is the same as that of Kulsrud {\sl et al.}

The assumption of a homogeneous magnetic field is warranted in the study
of rf heating and current drive in tokamaks if the rf interacts only
with circulating particles.  This is often not the case (for example
during ion- and electron-cyclotron heating) in which case proper
account should be taken of trapped particles.  This has been done in
so-called {\sl bounce-averaged} codes \cite{Cutler,Kerbel,Matsuda} in
which the distribution function is averaged over the bounce motion of
the trapped particles.  In tokamaks this leads to a modification of the
coefficients appearing in the Fokker--Planck equation but the numerical
treatment of the equation is largely unaltered.  In machines with more
complicated particle orbits, the distribution may be a multivalued
function of the velocity coordinates.  This occurs in tandem mirrors
where there is more than one population of trapped particles.  In this
case, special techniques are required \cite{Matsuda}.

\section{Preliminaries}\label{prelim}

\subsection{The Fokker--Planck equation}

We write the Fokker--Planck equation for the electrons $e$ as
	$${\partial f_e\over\partial t}-\sum_s C(f_e,f_s) +
\nabla\cdot{\bf S}_w+{q_e{\bf E}\over m_e}\cdot\nabla f_e = 0, \Eqlab{fp-1}$$
	where $q_s$ and $m_s$ are the charge and mass of species $s$,
$C(f_a,f_b)$ is the collision term for species $a$ colliding off
species $b$, the sum extends over all the species of the plasma
(typically electrons and ions), ${\bf S}_w$ is the wave ($w$)-induced
quasilinear flux, and ${\bf E} =E{\bf \hat v}_\parallel$ is the
electric field (assumed to be parallel to the magnetic field).  The
quantity $q_s$ carries the sign of the charge, thus $q_e=-e$.  The
subscripts $\parallel$ and $\perp$ refer to the directions parallel and
perpendicular to the magnetic field.  The
$\nabla\equiv\partial/\partial{\bf v}$ operator operates in velocity
space.

Because collisions in a plasma are primarily due to small-angle
scattering, the collision term can be written as the divergence of a
flux
	$$C(f_a,f_b)=-\nabla\cdot {\bf S}_c^{a/b},$$
in which case \eqref{fp-1} can be expressed as
	$${\partial f_e\over\partial t}+
\nabla\cdot{\bf S}= 0, \Eqlab{fp-2}$$
where 
	$${\bf S}={\bf S}_c+{\bf S}_w+
{\bf S}_e$$
	is the total flux in velocity space, and
$$\eqalignno{{\bf S}_c&=\sum_s{\bf S}_c^{e/s},&(\eqlab{s-coll})\cr
{\bf S}_e&={q_e{\bf E}\over m_e}f_e,&(\eqlab{s-e})\cr}$$
are the collisional ($c$)- and electric-field ($e$)-induced electron fluxes.

From \eqref{fp-2} we can derive the conservation laws
	$$\eqalignno{
{\partial\over\partial t}
\int_V f_e\,\dv+\int_A {\bf S}\cdot \da&=0,&(\eqlab{cons}a)\cr
{\partial\over\partial t}
\int_V m_e{\bf v}f_e\,\dv+
\int_A m_e{\bf v}{\bf S}\cdot \da&=\int_V m_e{\bf S}\,\dv,
&(\ref{cons}b)\cr
{\partial\over\partial t}
\int_V {m_ev^2\over2} f_e\,\dv+
\int_A {m_ev^2\over2}{\bf S}\cdot \da&=\int_V m_e{\bf v\cdot
S}\,\dv,&(\ref{cons}c)\cr}$$
	where $V$ is some volume in velocity space and $A$ is its
boundary.  These equations are statements of conservation of number,
momentum, and energy.

Typically, two types of terms appear in ${\bf S}$:  a diffusion term and
a friction term
	$${\bf S}=-\mat D\cdot \nabla f_e+ {\bf F} f_e.\Eqlab{diff-frict}$$
	The wave term is purely diffusive so that ${\bf F}_w=0$, while
the electric field term is nondiffusive: $\mat D_e=0$,
	$${\bf F}_e={q_e{\bf E}\over m_e}.\Eqlab{e-flux}$$

\subsection{Coordinate systems}

Because of azimuthal symmetry, $f_e$ is independent of $\phi$ the angle
about the magnetic field.  Two coordinate systems suggest themselves:
the cylindrical coordinate system $(v_\perp,v_\parallel,\phi)$ and the
spherical coordinate system $(v,\theta,\phi)$; see \figref{coord-fig}.
These are related by
	$$\eqalign{v^2&=v_\perp^2+v_\parallel^2,\cr
\cos\theta&=v_\parallel/v.\cr}$$
	Both of these coordinate systems are useful.  In cylindrical
coordinates (assuming azimuthal symmetry) \eqref{diff-frict} gives
	$$\eqalignno{\nabla\cdot{\bf S}&=
{1\over v_\perp}{\partial\over\partial v_\perp}v_\perp S_\perp
+{\partial\over\partial v_\parallel}S_\parallel,&(\eqlab{cylindrical}a)\cr
S_\perp&=
-D_{\perp\perp} {\partial f_e\over\partial v_\perp}
-D_{\perp\parallel}{\partial f_e\over\partial v_\parallel}
+F_{\perp}f_e,&(\ref{cylindrical}b)\cr
S_\parallel&=
-D_{\parallel\perp} {\partial f_e\over\partial v_\perp}
-D_{\parallel\parallel}{\partial f_e\over\partial v_\parallel}
+F_{\parallel}f_e.&(\ref{cylindrical}c)\cr}$$
	Similarly, in spherical coordinates we have
	$$\eqalignno{\nabla\cdot{\bf S}&=
{1\over v^2}{\partial\over\partial v}v^2S_v+
{1\over v\sin\theta}{\partial\over\partial \theta}\sin\theta S_\theta,
&(\eqlab{spherical}a)\cr
S_v&=
-D_{v v} {\partial f_e\over\partial v}
-D_{v \theta}{1\over v}{\partial f_e\over\partial \theta}
+F_{v}f_e,&(\ref{spherical}b)\cr
S_\theta&=
-D_{\theta v} {\partial f_e\over\partial v}
-D_{\theta \theta}{1\over v}{\partial f_e\over\partial \theta}
+F_{\theta}f_e.&(\ref{spherical}c)\cr}$$
	Transformations between $\mat D$ and $\bf F$ expressed in the
two coordinate systems may be achieved by
	$$\pmatrix{D_{\perp\perp}\cr D_{\perp\parallel}\cr
D_{\parallel\perp}\cr D_{\parallel\parallel}\cr}
=\mat M\cdot \pmatrix{D_{v v}\cr D_{v \theta}\cr
D_{\theta v}\cr D_{\theta \theta}\cr}
\eqn(\eqlab{transform}a)$$
	and
	$$\pmatrix{F_{\perp}\cr F_{\parallel}\cr}=
\mat N\cdot \pmatrix{F_{v}\cr F_{\theta}\cr},
\eqn(\ref{transform}b)$$
	where
	$$\eqalign{\mat M=\mat M^{-1}&=
\pmatrix{s^2&sc&sc&c^2\cr   sc&-s^2&c^2&-sc\cr
         sc&c^2&-s^2&-sc\cr c^2&-sc&-sc&s^2\cr}\cr
\mat N=\mat N^{-1}&=
\pmatrix{s&c\cr c&-s\cr},\cr}$$
	and here we have abbreviated $s=\sin\theta$ and $c=\cos\theta$.
The collision term is most conveniently expressed in spherical
coordinates and \eqsref{transform} allow us to transform this term to
cylindrical coordinates.  On the other hand, the rf and electric field
terms are written most naturally in cylindrical coordinates, and this
equation also enables us to express these terms in spherical
coordinates.

In the case of the collision operator, $\mat D$ and $\bf F$ are given
in terms of the gradients of potentials
$$\mat D \propto \nabla\nabla\psi,\qquad {\bf F}\propto\nabla\phi.$$
	In cylindrical coordinates (with azimuthal symmetry), the
relevant components of $\mat D$ and $\bf F$ are easy to calculate---we
just take the corresponding derivatives of $\psi$ and $\phi$.  In
spherical coordinates we have
	$$\eqalignno{(\nabla\nabla\psi)_{vv}&=
{\partial^2 \psi\over\partial v^2},
&(\eqlab{spherical-comp}a)\cr
(\nabla\nabla\psi)_{v\theta}=(\nabla\nabla\psi)_{\theta v}&=
{1\over v}{\partial^2 \psi\over\partial v\partial \theta}
-{1\over v^2}{\partial \psi\over\partial \theta},&(\ref{spherical-comp}b)\cr
(\nabla\nabla\psi)_{\theta\theta}&={1\over v}{\partial \psi\over\partial v}
+{1\over v^2}{\partial^2 \psi\over\partial \theta^2},&(\ref{spherical-comp}c)\cr
(\nabla\phi)_v&={\partial \phi\over\partial v},&(\ref{spherical-comp}d)\cr
(\nabla\phi)_\theta&={1\over v}{\partial \phi\over\partial \theta}.
&(\ref{spherical-comp}e)\cr}$$

Several important quantities are given in terms of velocity-space
moments of the distribution function.  Three-dimensional velocity
space integrations can be carried out in cylindrical coordinates using
	$$\int f({\bf v})\,\dv
=\int_0^\infty\! dv_\perp\int_{-\infty}^\infty\! dv_\parallel
\, 2\pi v_\perp f(v_\perp,v_\parallel)
\Eqlab{int-cyl}$$
	and in spherical coordinates using
	$$\int f({\bf v})\,\dv
=\int_0^\infty\! dv\int_0^\pi\! d\theta\, 2\pi v^2 f(v,\theta)\sin\theta.
\Eqlab{int-spher}$$

\subsection{Legendre harmonics}

	It is sometimes useful to decompose the distribution function
and the potentials into Legendre harmonics $P_l(\cos\theta)$.  We write
this as
	$$f(v,\theta)=\sum_{l=0}^\infty f^{(l)}(v)P_l(\cos\theta),
\Eqlab{legend-comp}$$
where
$$f^{(l)}(v)={2l+1\over 2}
\int_0^\pi f(v,\theta) P_l(\cos\theta)\sin\theta\,d\theta.
\Eqlab{legend-decomp}$$
	The Legendre polynomials may be evaluated on a computer using
the recurrence relation \cite{Abramowitz}
	$$P_0(\mu)=1,\qquad
	(l+1)P_{l+1}(\mu)=(2l+1)\mu P_l(\mu)-lP_{l-1}(\mu).$$

\subsection{Definitions}

Finally, we define some of the other quantities that we encounter in
this paper.  The thermal velocity of species $s$ is given by
$$v_{ts}=\sqrt{T_s\over m_s},\Eqlab{vts-max}$$
	where $T_s$ is the temperature of species $s$.  The thermal
collision frequency for the electrons is
	$$\nu_{te}=\tau_{te}^{-1}={\Gamma^{e/e}\over 
v_{te}^3},\Eqlab{nute}$$
where
$$\Gamma^{a/b}={n_b q_a^2 q_b^2 \ln \Lambda^{a/b}\over 4\pi
\epsilon_0^2 m_a^2},$$
	and $n_s$ is the number density of species $s$, $\epsilon_0$ is
the dielectric constant of free space, and $\ln \Lambda^{a/b}$ is the
Coulomb logarithm.  The distributions are normalized so that
	$$\int f_s({\bf v})\,\dv=n_s.$$
	In particular, the Maxwellian distribution is
$$\eqalignno{f_{sm}(v)&=n_s\biggl({m_s\over 2\pi T_s}\biggr)^{3/2}
 \exp(-\half m_sv^2/T_s),\cr
&=n_s{1\over(2\pi v_{ts}^2)^{3/2}}
 \exp(-\half v^2/v_{ts}^2).&
(\eqlab{max})\cr}$$

In discussing the applications to rf current drive, there are two
quantities in which we will be interested: the electron current density
	$$J=\int q_ev_\parallel f_e({\bf v})\,\dv\Eqlab{curr-def}$$
and the rf power absorbed per unit volume by the plasma
	$$P=\int m_e{\bf v}\cdot{\bf S}_w\,\dv.\Eqlab{pd-def}$$
	The efficiency of rf current drive is usually given as the
ratio $J/P$.

We shall use {\sc S.I.} units throughout this paper except that we will
measure temperature in units of energy.

\section{Collision Operator}\label{coll}
\subsection{The Landau collision operator}

The collision flux is given by the Landau collision integral
\cite{Landau}
	$${\bf S}_c^{a/b}={q_a^2 q_b^2 \over 8\pi\epsilon_0^2 m_a}
\ln \Lambda^{a/b}\int \mat U({\bf u})\cdot
\biggl(
{f_a({\bf v})\over m_b} {\partial f_b({\bf v}')\over\partial {\bf v}'}-
{f_b({\bf v}')\over m_a} {\partial f_a({\bf v})\over\partial {\bf v}}
\biggr) \,\dv',\Eqlab{coll-land}$$
	where
	$$
\mat U({\bf u})={u^2\mat I - {\bf uu}\over u^3},\qquad
{\bf u}={\bf v}-{\bf v}'.$$
	The formula for the Coulomb logarithm $\ln \Lambda^{a/b}$
is given in text books and the {\sl NRL Plasma Formulary} \cite{NRL}.  Because it
is so insensitive to plasma parameters, in many cases it is adequate to
take it to be a constant equal to $15$.  In any case, it is required
that $\ln\Lambda^{a/b}=\ln\Lambda^{b/a}$.  The Landau collision operator
conserves number, momentum, and energy, i.e.,
	$$\eqalignno{\int C(f_a,f_b)\,\dv&=0,
&(\eqlab{land-cons}a)\cr
\int m_a{\bf v}C(f_a,f_b)\,\dv+
\int m_b{\bf v}C(f_b,f_a)\,\dv&=0,&(\ref{land-cons}b)\cr
\int {m_a v^2\over2}C(f_a,f_b)\,\dv+
\int {m_b v^2\over2}C(f_b,f_a)\,\dv&=0.&(\ref{land-cons}c)\cr}$$

There is an error on the order of $1/\ln\Lambda^{a/b}$ in the Landau
collision operator.  However, because it has so many ``nice''
properties---the conservation laws of \eqsref{land-cons}, an
$H$-theorem, etc.---it is customary to regard \eqref{coll-land} as
being exact.

\subsection{Rosenbluth potentials}

\Eqref{coll-land} is the most useful form for the collision operator
for analytical work.  However, it is not in a convenient form for
numerical computations.  Suppose we represent the distribution functions
on an $N\times N$ grid (assuming azimuthal symmetry).  Then evaluation of
\eqref{coll-land} entails $O(N^4)$ computations, because it entails a
two-dimensional integral (over ${\bf v}'$) which must be carried out at
each grid location.  Fortunately, substantial savings may be realized by
using an equivalent representation in terms of Rosenbluth potentials
\cite{Rosenbluth,Trubnikov1}.  Here we use the slightly more convenient
notation of Trubnikov \cite{Trubnikov}.  We define two potentials
	$$\eqalignno{
\phi_b({\bf v})&=-{1\over 4\pi}
\int {f_b({\bf v}')\over\abs{{\bf v}-{\bf v}'}}\,\dv',
&(\eqlab{pot}a)\cr
\psi_b({\bf v})&=-{1\over 8\pi}
\int \abs{{\bf v}-{\bf v}'}f_b({\bf v}')\,\dv'.
&(\ref{pot}b)\cr}$$
	These are called potentials because they
satisfy Poisson's equations in velocity space
$$\nabla^2\phi_b({\bf v})=f_b({\bf v}),\qquad
\nabla^2\psi_b({\bf v})=\phi_b({\bf v}).$$
	In terms of these potentials, \eqref{coll-land} becomes
	$$\eqalignno{{\bf S}_c^{a/b}&=
-\mat D_c^{a/b}\nabla f_a({\bf v})+{\bf F}_c^{a/b}f_a({\bf v}),&
(\eqlab{coll-ros}a)\cr
\mat D_c^{a/b}&=-{4\pi\Gamma^{a/b}\over n_b}
\nabla\nabla\psi_b({\bf v}),&(\ref{coll-ros}b)\cr
{\bf F}_c^{a/b}&=-{4\pi\Gamma^{a/b}\over n_b}
{m_a\over m_b}\nabla\phi_b({\bf v}).&(\ref{coll-ros}c)\cr}$$
	(An equivalent form of this equation is given by Rosenbluth
{\sl et al.}\ \cite{Rosenbluth} which contains a term of the form
	$$\nabla\cdot[f_a({\bf v})\nabla\nabla\psi_b({\bf v})].$$
	This form is used in several numerical codes
\cite{Killeen1,Killeen2,McCoy}, even though more derivatives of
$\psi_b$ must be taken.  One form may be derived from the other by
noting that $\nabla^2\psi_b=\phi_b$.)

	There is an efficient method for calculating the Rosenbluth
potentials.  This involves decomposing $f_b$ in Legendre harmonics
\eqref{legend-decomp}.  Then we have \cite{Rosenbluth}
	$$\eqalignno{\phi_b^{(l)}(v)&=-{1\over 2l+1}
\biggl[\int_0^v {v'{}^{l+2}\over v^{l+1}} f_b^{(l)}(v')\,dv'
+\int_v^\infty {v^l\over v'{}^{l-1}}f_b^{(l)}(v')\,dv'\biggr],&
(\eqlab{pot1}a)\cr
\psi_b^{(l)}(v)&={1\over 2(4l^2-1)}
\biggl[\int_0^v {v'{}^{l+2}\over v^{l-1}}
\biggl(1-{l-\half\over l+\fract3/2}{v'{}^2\over v^2}\biggr)
f_b^{(l)}(v')\,dv'\cr&\qquad\qquad\qquad
+\int_v^\infty {v^l\over v'{}^{l-3}}
\biggl(1-{l-\half\over l+\fract3/2}{v^2\over v'{}^2}\biggr)
f_b^{(l)}(v')\,dv'\biggr].&
(\ref{pot1}b)\cr}$$
	Let us assume that $f_b$ may be represented by $K$ Legendre
harmonics (i.e., the upper limit in the sum in \eqref{legend-comp} is
$K-1$).  Then the calculation of $f_b^{(l)}(v)$ from $f_b({\bf v})$
using \eqref{legend-decomp} takes $O(N^2)$ computations for each $l$ or
$O(KN^2)$ computations altogether.  Given $f_b^{(l)}(v)$, the
calculation of $\phi_b^{(l)}(v)$ and $\psi_b^{(l)}(v)$ using
\eqsref{pot1} takes $O(N)$ computations for each $l$.  The
calculation of $\phi({\bf v})$ and $\psi({\bf v})$ takes a further
$O(KN^2)$ step.  Overall the number of steps is therefore $O(KN^2)$.
Often, we can take $K$ to be quite small (usually $K<10$), and in any
case we have $K\le N$, so we can compute the collision term much more
economically than using the Landau operator directly.

\section{Approximations to the Collision Operator}\label{coll-approx}
\subsection{Isotropic background}

If the background distribution is isotropic $f_b({\bf v})=f_b(v)$, then
so too are $\phi$ and $\psi$.  The collision term is then from
\eqsref{coll-ros}, (\ref{pot1}), and (\ref{spherical-comp})
	$$\eqalignno{S_{cv}^{a/b}&=
-D_{cv v}^{a/b} {\partial f_a\over\partial v}
+F_{cv}^{a/b}f_a,&(\eqlab{isotropic}a)\cr
S_{c\theta}^{a/b}&=
-D_{c\theta \theta}^{a/b}{1\over v}{\partial f_a\over\partial \theta}
,&(\ref{isotropic}b)\cr}$$
	where
	$$\eqalignno{D_{cvv}^{a/b}&=
{4\pi\Gamma^{a/b}\over 3n_b}
\biggl(\int_0^v {v'{}^4\over v^3}f_b(v')\,dv'
+\int_v^\infty v' f_b(v')\,dv'\biggr),&(\eqlab{iso-gen}a)\cr
D_{c\theta\theta}^{a/b}&=
{4\pi\Gamma^{a/b}\over 3n_b}
\biggl(\int_0^v {v'{}^2\over 2v^3}(3v^2-v'{}^2)f_b(v')\,dv'
+\int_v^\infty v' f_b(v')\,dv'\biggr),&(\ref{iso-gen}b)\cr
F_{cv}^{a/b}&=
-{4\pi\Gamma^{a/b}\over 3n_b}
{m_a\over m_b}
\int_0^v{3v'{}^2\over v^2}f_b(v')\,dv'.&(\ref{iso-gen}c)\cr}$$

\subsection{The high-velocity limit}

If $v$ is much greater than the thermal velocity of particles of
species $b$, the indefinite limits in \eqsref{iso-gen} may be replaced
by infinity to give
	$$\eqalignno{D_{cvv}^{a/b}&=
\Gamma^{a/b} {v_{tb}^2\over v^3}
,&(\eqlab{iso-high}a)\cr
D_{c\theta\theta}^{a/b}&=
\Gamma^{a/b} {1\over 2v}\biggl(1-{v_{tb}^2\over v^2}\biggr)
,&(\ref{iso-high}b)\cr
F_{cv}^{a/b}&=
-\Gamma^{a/b} {m_a\over m_b}
{1\over v^2},&(\ref{iso-high}c)\cr}$$
	where the thermal velocity is defined for an arbitrary isotropic
distribution as
	$$v_{ts}^2={4\pi\over 3n_s}\int_0^\infty v^4 f_s(v)\,dv.
\Eqlab{vts}$$
[For a Maxwellian distribution, \eqref{max}, this reduces to the usual
expression \eqref{vts-max}.]

\subsection{Maxwellian background}

If the background distribution is Maxwellian \eqref{max}, the integrals
in \eqref{iso-gen} can be carried out to give \cite{Trubnikov}
\def\erf{\mathop{\rm erf}\nolimits}%
	$$\eqalignno{D_{cvv}^{a/b}&={\nu_\parallel^{a/b}\over 2}v^2
={\Gamma^{a/b}\over 2v}\biggl({\erf(u)\over u^2}-{\erf'(u)\over u}
\biggr),&(\eqlab{iso-max}a)\cr
D_{c\theta\theta}^{a/b}&={\nu_\perp^{a/b}\over 4}v^2
={\Gamma^{a/b}\over 4v}\biggl(\biggl(2-{1\over u^2}\biggr)\erf(u)
+{\erf'(u)\over u}\biggr),&(\ref{iso-max}b)\cr
F_{cv}^{a/b}&=-{m_a\over m_a+m_b}\nu_s^{a/b}v
=-{\Gamma^{a/b}\over v^2}{m_a\over m_b}\bigl[\erf(u)-u\erf'(u)
\bigr],&(\ref{iso-max}c)\cr}$$
	where
$$\eqalign{\erf(u)&={2\over\sqrt\pi}\int_0^u\exp(-x^2)\,dx,\cr
\erf'(u)&={2\over\sqrt\pi}\exp(-u^2),\cr
u&={v\over\sqrt2 v_{tb}}.\cr}$$
	The parallel diffusion rate $\nu_\parallel^{a/b}$,
perpendicular diffusion rate $\nu_\perp^{a/b}$, and the slowing down
diffusion rate $\nu_s^{a/b}$ are the same as those defined in the {\sl
NRL Plasma Formulary} \cite{NRL}.  [However, the {\sl NRL Plasma
Formulary} (1983 edition) has an incorrect formula for the collision
operator with a Maxwellian background.]

	For $u>0$, $\erf(u)$ is given approximately by \cite{Abramowitz}
	$$\erf(u)=1-\exp(-u^2)\sum_{k=1}^5 a_k t^k,$$
	where
	$$t=1/(1+pu)$$ and
	$$\eqalign{p&=0.32759\,11,\cr
	a_1&=0.25482\,9592,\qquad a_2=-0.28449\,6736,\cr
	a_3&=1.42141\,3741,\qquad a_4=-1.45315\,2027,\cr
	a_5&=1.06140\,5429.\cr}$$
	This approximation cannot be used in evaluating
\eqsref{iso-max} near $u=0$ because there is cancellation to leading
order in all three terms.  In that case, the Taylor expansion,
	$$\eqalign{\erf(u)-u\erf'(u)&={2\over\sqrt\pi}
\sum_{k=0}^\infty {(-1)^k\over k!}{2\over 2k+3}u^{2k+3}
\cr&={2\over\sqrt\pi}\biggl({2\over 3}u^3 - {2\over 5}u^5+{1\over 7}u^7-{1\over27}
u^9+\ldots\biggr),\cr}$$
	may be used.  Alternatively, we can compute \eqsref{iso-max} by
numerically evaluating the integrals in \eqsref{iso-gen} with
$f_b(v)=f_{bm}(v)$.  This method is then easily extended to include
relativistic effects as given in \secref{relativity}.

The ratio $F_{cv}^{a/b}/D_{cvv}^{a/b}$ from \eqsref{iso-max} [and also
(\ref{iso-high})] is $-m_av/T_b$ so that the effect of collisions with
species $b$ is to make species $a$ approach a Maxwellian with
temperature $T_b$.

\subsection{Linearized collision operator}

In many applications in plasma physics (including those involving rf
waves) collisions dominate the thermal particles.  Therefore, the
distribution function may be expanded about a Maxwellian
	$$f_a({\bf v})=f_{am}(v)+f_{a1}({\bf v}).$$
The self-collision operator $C(f_a,f_a)$ may be approximated by the
linearized operator
	$$
C_{\rm lin}^{a/a}\bigl(f_a({\bf v})\bigr)=
	C\bigl(f_a({\bf v}),f_{am}(v)\bigr)+
C\bigl(f_{am}(v),f_a({\bf v})\bigr),
\Eqlab{linear}$$
	where we have made use of the fact that $C(f_{am},f_{am})=0$,
and we have ignored terms of order $f_{a1}^2$.  We can compute
$C\bigl(f_a({\bf v}),f_{am}(v)\bigr)$ using \eqsref{isotropic} and
(\ref{iso-max}).  To compute $C\bigl(f_{am}(v),f_a({\bf v})\bigr)$, we
express $f_a({\bf v})$ as a sum of Legendre harmonics,
\eqref{legend-comp}, to give
	$$C\bigl(f_{am}(v),f_a({\bf v})\bigr)=\sum_{l=0}^\infty
C\bigl(f_{am}(v),f_a^{(l)}({\bf v})P_l(\cos\theta)\bigr).
\Eqlab{linear-expand}$$
	The zeroth term in the sum can be computed using
\eqsref{isotropic} and (\ref{iso-gen}) giving
	$$\eqalignno{{C\bigl(f_{am}(v),f_a^{(0)}(v)\bigr)\over f_{am}(v)}
={4\pi\Gamma^{a/a}\over n_a}\biggl[
f_a^{(0)}(v)
&{}+\int_0^v {v'{}^2\over v_{ta}^2}f_a^{(0)}(v')
\biggl({v'{}^2\over 3v_{ta}^2v}-{1\over v}\biggr)\,dv'\cr
&{}+\int_v^\infty {v'{}^2\over v_{ta}^2}f_a^{(0)}(v')
\biggl({v^2\over 3v_{ta}^2v'}-{1\over v'}\biggr)\,dv'\biggr].
&(\eqlab{iso-lin})\cr}$$

The next term in the sum in \eqref{linear-expand} is
given by \eqsref{coll-ros}, (\ref{pot1}), and (\ref{spherical-comp})
	$$\eqalignno{
{C\bigl(f_{am}(v), f_a^{(1)}(v)\cos\theta\bigr)\over f_{am}(v)\cos\theta}=
{4\pi\Gamma^{a/a}\over n_a}\biggl[f_a^{(1)}(v)
&{}+\int_0^v {v'{}^2\over v_{ta}^2}f_a^{(1)}(v')
\biggl({v'{}^3\over 5v_{ta}^2v^2}-{v'\over 3v^2}\biggr)\,dv'\cr
&{}+\int_v^\infty {v'{}^2\over v_{ta}^2}f_a^{(1)}(v')
\biggl({v^3\over 5v_{ta}^2v'{}^2}-{v\over 3v'{}^2}\biggr)\,dv'\biggr].
&(\eqlab{first-lin})\cr
}$$
The corresponding fluxes for this term are
$$\eqalignno{{S_v^{a/a}\over f_{am}(v)\cos\theta}=
{4\pi\Gamma^{a/a}\over n_a}\biggl[
&\int_0^v f_a^{(1)}(v')
\biggl({v'{}^5\over 5v_{ta}^2v^3}-{2v'{}^3\over 3v^3}\biggr)\,dv'\cr
&+\int_v^\infty f_a^{(1)}(v')
\biggl({v^2\over 5v_{ta}^2}+{1\over 3}\biggr)\,dv'\biggr],
&(\eqlab{first-flux}a)\cr
{S_\theta^{a/a}\over f_{am}(v)\sin\theta}=
{4\pi\Gamma^{a/a}\over n_a}\biggl[
&\int_0^v f_a^{(1)}(v')
\biggl({v'{}^5\over 10v_{ta}^2v^3}-{v'{}^3\over 6v_{ta}^2v}
-{v'{}^3\over 3v^3}\biggr)\,dv'\cr
&-\int_v^\infty f_a^{(1)}(v')
\biggl({v^2\over 15v_{ta}^2}+{1\over
3}\biggr)\,dv'\biggr].&(\ref{first-flux}b)\cr
}$$

Because of conservation of number, momentum, and energy, the solutions
to the homogeneous equation
	$$
C_{\rm lin}^{a/a}\bigl(f_a({\bf v})\bigr)=0$$
are
	$$f_a({\bf v})=(a+{\bf b}\cdot m_a{\bf v}+d\half m_av^2)
f_{am}(v).$$
	If we substitute $a={\bf b}=0$, we obtain a check on
\eqref{iso-lin}. Similarly, $a=d=0$ and ${\bf b}={\bf \hat v}_\parallel$
gives a check on \eqref{first-lin}.  Such checks are useful when
incorporating the linearized collision operator into a numerical code.

\subsection{Electron-ion collision operator}

We now turn to the specific problem of current drive by lower hybrid
waves.  In this problem we wish to solve the Fokker--Planck equation for
the electrons including the effects of electron-ion and
electron-electron collisions.

Because the ions are so massive relative
to the electrons, we have $v\gg v_{ti}$ for nearly all the electrons and
\eqsref{iso-high} apply.  Indeed, we can make the further approximations
$m_i\rightarrow\infty$, $v_{ti}\rightarrow 0$, in which case the
collision operator is given by \eqref{isotropic} with
	$$\eqalignno{D_{cvv}^{e/i}&=F_{cv}^{e/i}=0
,&(\eqlab{ions}a)\cr
D_{c\theta\theta}^{e/i}&=
\Gamma^{e/e}{Z_i\over 2v},&(\ref{ions}b)\cr}$$
	where
	$$Z_i=-{q_i\ln\Lambda^{e/i}\over q_e\ln\Lambda^{e/e}},$$
	and we have assumed neutrality $q_en_e+q_in_i=0$.  The full
electron-ion collision term $C(f_e,f_i)$ can be written as
	$$C^{e/i}\bigl(f_e({\bf v})\bigr)=
\Gamma^{e/e}{Z_i\over 2v^3}{1\over\sin\theta}{\partial\over\partial\theta}
\sin\theta{\partial\over\partial\theta}f_e({\bf v}).\Eqlab{coll-ions}$$
For a
multispecies plasma $Z_i$ is replaced by $Z_{\rm eff}$ where
	$$Z_{\rm eff}=
{\sum_s n_s q_s^2\ln\Lambda^{e/s}\over n_e q_e^2\ln\Lambda^{e/e}},$$
and the sum extends over all the ionic species.

With this collision operator the ions are characterized by a single
dimensionless parameter $Z_i$ (or $Z_{\rm eff}$).  The collision
operator allows momentum to be transferred from the electrons to the
ions, but there is no energy exchange.  The non-negative nature of
$f_e$ is preserved.

\subsection{Electron-electron collision operator}

There are several choices for the electron-electron collision
operator. We will discuss them starting at the most complex.

The {\sl full} electron-electron collision operator is given by
\eqsref{coll-ros} and (\ref{pot1}).  This was first used in
current-drive studies by Harvey {\sl et al.}\ \cite{Harvey}.  This
collision operator conserves both momentum and energy.  The electron
distribution $f_e$ remains non-negative.  Because the collision operator
conserves energy, there is nowhere for rf energy absorbed by the
electrons to go.  The problem arises because we have reduced a problem
in configuration and velocity space to one in velocity space alone, so
that there is now no spatial diffusion of energy.  In practice, this
problem is solved by inserting an energy loss term into the
Fokker--Planck \eqref{fp-1}.  Unfortunately, there are several
different models for this loss term and so this procedure is somewhat
{\sl ad hoc}.

The {\sl linearized} electron-electron collision operator is
given by \eqref{linear}.  This too conserves momentum and energy.
The non-negative nature of $f_e$ is no longer preserved.  
When the perturbation to $f_{em}$ is small, $f_e$ usually only
becomes negative far out on the tail.  The energy conservation of the
collision operator again necessitates the introduction of an energy
loss term.  Fortunately, there is a systematic way to do this within
the context of a Chapman--Enskog--Braginskii expansion
\cite{Chapman,Braginskii}.  The energy loss term has the form
	$$-\biggl({m_ev^2\over 2T_e}-{3\over 2}\biggr)f_{em}(v)
{\partial\ln T_e\over \partial t},$$
	which appears on the right-hand side of \eqref{fp-1}.
Operationally, ${\partial\ln T_e/ \partial t}$ would be adjusted to
ensure that the energy of the electron distribution $f_{e}({\bf v})$
remained constant.  (In fact, one of the results of the expansion
procedure is an equation for the evolution of $T_e$ including the
effects of rf and ohmic heating and of energy transport.)

A useful modification of this collision operator is the {\sl truncated}
collision operator
	$$C_{\rm trunc}^{e/e}\bigl(f_e({\bf v})\bigr)=
C\bigl(f_e({\bf v}),f_{em}(v)\bigr)
+C\bigl(f_{em}(v), f_e^{(1)}(v)\cos\theta\bigr),
\Eqlab{truncated}$$
	where the first term is given by \eqsref{isotropic} and
(\ref{iso-max}) and the second term by \eqref{first-lin}.  This differs
from the linearized operator in that we only retain the $l=0$ and $l=1$
terms in the sum in \eqref{linear-expand} and we further approximate
$f_e^{(0)}(v)$ by $f_{em}(v)$.  As a consequence, this operator
conserves momentum but not energy; so there is no need to introduce an
energy loss term.  Again, the electron distribution function may become
negative.  This collision operator is useful in the study of current
drive by low-phase-velocity waves and in the treatment of problems with
an electric field.  In both of these examples, a momentum-conserving
electron-electron collision operator is required.  This operator was
used (in a relativistic form) in the study of current drive by fast
waves \cite{relpap}.

A slightly different technique for ensuring momentum conservation  was
used in our study of current drive by low-phase-velocity waves
\cite{Fisch-Karney}.  There we approximated the electron-electron
collision operator by
$$C_{\rm drift}^{e/e}\bigl(f_e({\bf v})\bigr)=
C\bigl(f_e({\bf v}),f_{em}(\abs{{\bf v}-v_d{\bf \hat v}_\parallel})\bigr),$$
	where the background is a {\sl drifting} Maxwellian with a drift
speed $v_d$ adjusted so that the parallel force between $f_e({\bf v})$
and the drifting Maxwellian,
	$$P_\parallel^{e/e}=\int m_e S_\parallel^{e/e} \,\dv,
\Eqlab{force-def}$$
	vanishes.  This collision operator conserves momentum (by
construction) and preserves the non-negative nature of $f_e$.  Energy is
not conserved.  It is, however, slightly less accurate than the truncated
operator.  In particular, while the truncated operator gives the correct
value for the electrical conductivity \cite{Spitzer}, this operator
gives an answer which is in error by about $15\%$.  The computation of
this collision operator involves computing \eqsref{iso-max} in the
drifting frame, converting to cylindrical coordinates using
\eqsref{transform}, transforming to the rest frame (which is easy in
cylindrical coordinates), and finally converting back to spherical
coordinates using \eqsref{transform}.  In order to
determine the drift speed, we use the analytical formula for the force
on an electron Maxwellian drifting with speed $v_d$ due to a stationary
ion background, i.e.,
	$$P_\parallel^{e/i}=-n_e m_e \nu_{te} v_d Z_i
{1\over 3}\sqrt{2\over \pi}\sqrt{m_i\over m_i+m_e},$$
	which is valid for $\abs{v_d}\ll v_{te}$.  Here we have taken
the mass ratio $m_e/m_i$ to be finite and have assumed that $T_i=T_e$.
The force between two electron Maxwellians with a relative drift of
$v_d$ is found by taking $Z_i=1$ and $m_i=m_e$ which gives
	$$P_\parallel^{e/e}=-{n_e m_e \nu_{te} v_d
\over 3\sqrt\pi}.\Eqlab{force-max}$$
	In the numerical code $P_\parallel^{e/e}$ is computed using
\eqref{force-def}.  \Eqref{force-max} is used to estimate the {\sl change} in
$v_d$ required to give $P_\parallel^{e/e}=0$.

The situation may be further simplified by assuming that the
background electrons are {\sl Maxwellian}, so that the collision
operator is given by
	$$C_{\rm Max}^{e/e}\bigl(f_e({\bf v})\bigr)=
C\bigl(f_e({\bf v}),f_{em}(v)\bigr),
\Eqlab{coll-max}$$
	which may be evaluated using \eqsref{isotropic} and
(\ref{iso-max}).  This operator conserves neither energy nor momentum.
It does preserve the non-negative nature of $f_e$.  It was used by
Kulsrud {\sl et al.}\ \cite{Kulsrud} in the study of runaways, and in
studies of lower hybrid current drive \cite{Karney-lh}.  The Maxwellian
background serves as a heat bath, so no energy loss terms are required.
The absence of momentum conservation introduces approximately a
factor-of-two error in the electrical conductivity \cite{Kulsrud} and
in the efficiency of current drive by slow waves \cite{Fisch-Karney}.
There is a relative error of order $(v_{te}/v)^3$ in the determination
of the current-drive efficiency for fast waves \cite{relpap}.

Lastly, $C_{\rm Max}^{e/e}$ may be approximated by using the {\sl high}
velocity limit, i.e., by using \eqsref{iso-high} instead of
\eqsref{iso-max}.  In fact, because eq.~(\ref{iso-high}b) gives negative
diffusion for small $v$, it is usually replaced by
	$$D_{c\theta\theta}^{a/b}=\Gamma^{a/b} {1\over 2v}.$$
	We define the resulting electron-electron collision operator as
$C_{\rm high}^{e/e}$.  It has much the same properties as $C_{\rm
Max}^{e/e}$.  In particular, it yields a Maxwellian (with
temperature $T_e$) as the steady-state solution.  Because of the greater
error in the collision term for thermal particles the electrical
conductivity is even lower than for $C_{\rm Max}^{e/e}$.  The evaluation
of \eqsref{iso-high} is, of course, a little easier to program than that
of \eqsref{iso-max}.  However, because the results of evaluating
\eqsref{iso-max} can be stored in a table, the extra computational cost of
working with $C_{\rm Max}^{e/e}$ is insignificant compared to the
solution of the Fokker--Planck equation.  Since $C_{\rm high}^{e/e}$ is
less accurate, its use is not recommended for numerical
work.  It is, however, useful in analytical work.

When working with these electron-electron collision operators, it is
useful to have some benchmark against which to check their numerical
realization.  A useful benchmark is provided by the electrical
conductivity, which is the ratio of electrical current to electric field
in the limit $E\rightarrow 0$.  This is tabulated in \tabref{conduct}
for various values of $Z_i$ and for all the electron-electron collision
operators discussed here.  These values were obtained by solving the
corresponding one-dimensional equation by the method outlined in
\secref{adjoint}.  In all cases, the electron-ion collision operator is
given by \eqref{coll-ions}.  The conductivity using the full and the truncated
electron-electron collision operators is the same as for the
linearized operator.
In the limit $Z_i\rightarrow\infty$, the conductivity is independent of
the electron-electron collision model
	$${J\over E}=16\sqrt{2\over\pi}{1\over Z_i}
{n_eq_e^2\over m_e\nu_{te}}.$$
	For the high-velocity approximation to the collision operator
the conductivity can be expressed analytically as
	$${J\over E}=
{16\over3}\sqrt{2\over\pi}{3Z_i+13\over(Z_i+3)(Z_i+5)}
{n_eq_e^2\over m_e\nu_{te}}.$$

\section{Quasilinear Operator}\label{ql}

\subsection{Single wave}

The interaction of electrons (or other species) with a wave is
conveniently described in terms of the quasilinear theory
\cite{Kennel}.  In this theory the flux of electrons in velocity space is
given by
	$${\bf S}_w=-\mat D_w\cdot\nabla f_e,\Eqlab{wave-flux}$$
	where $\mat D_w$ is the quasilinear diffusion tensor which
depends on the waves present in the plasma.  Although quasilinear theory
is not strictly applicable to a single wave, we will start with this
case because it is the simplest.  Suppose there is a uniform wave
present in the plasma, i.e., \def\re{\mathop{\rm Re}\nolimits}%
	$${\bf E}({\bf r},t)=\re[{\bf E}_w\exp(i{\bf k\cdot r}-i\omega t)].
\Eqlab{single-wave}$$
	The quasilinear diffusion coefficient is given by \cite{Kennel}
	$$\mat D_w=\sum_n{\pi\over 2}{q_e^2\over m_e^2}
\delta(\omega-k_\parallel v_\parallel-n\Omega_e){\bf a}_n^\ast{\bf a}_n
\Eqlab{ql-single}$$
and
$$\eqalignno{
{\bf a}_n&=\Theta_n{k_\parallel\over \omega}\biggl[\biggl(
{\omega\over k_\parallel}-v_\parallel\biggr){\bf \hat v}_\perp+
v_\perp{\bf \hat v}_\parallel\biggl]\cr
\Theta_n&={E_{w+} J_{n-1} + E_{w-} J_{n+1}\over \sqrt 2}+
{v_\parallel\over v_\perp}J_nE_{w\parallel},\cr}$$
	where $\Omega_e=q_eB/m_e$ is the electron cyclotron frequency,
$B$ is the magnetic field, $\ast$ indicates complex conjugation,
 $J_n$ is the $n$th order Bessel function,
and the argument of the Bessel functions is $k_\perp v_\perp/\Omega_e$.
$E_{w+}$ and $E_{w-}$ are the left- and right-handed components of
${\bf E}_w$; in a right-handed
cartesian coordinate system with $\bf \hat z$ parallel to $\bf B$ and
$\bf k$ lying in the $(x,z)$ plane, we
have
	$$\eqalign{E_{w+}&={E_{wx}+iE_{wy}\over\sqrt2},\cr
E_{w-}&={E_{wx}-iE_{wy}\over\sqrt2},\cr
E_{w\parallel}&=E_{wz}.\cr}$$

It is instructive to consider the properties of \eqref{ql-single}.  The
delta function specifies the resonance condition.  Only particles for
which the Doppler-shifted wave frequency $\omega-k_\parallel
v_\parallel$ is zero ($n=0$---the Landau resonance) or a multiple of
the cyclotron frequency ($n\ne 0$---a cyclotron harmonic resonance)
interact with the wave.  The vector ${\bf a}_n$ is perpendicular to the
velocity of the electron in the wave frame ${\bf
v}-(\omega/k_\parallel){\bf \hat v}_\parallel$.  This means that the
wave-induced flux is along diffusion paths which lie in constant-energy
surfaces in the wave frame; see \figref{diff-fig}.  Similarly, the flux
is proportional to the gradient in $f_e$ in this direction.  As a
consequence, when an electron interacts with a particle via the Landau
resonance, the diffusion tensor consists of only a single component
	$$\mat D_w=D_{\parallel\parallel}
{\bf \hat v}_\parallel{\bf \hat v}_\parallel.$$
Likewise, for a cyclotron harmonic resonance, we have
$$\mat D_w=D_{\perp\perp}
{\bf \hat v}_\perp{\bf \hat v}_\perp,$$
provided that $v_\perp$ is small compared with $n\Omega_e/k_\parallel$.

The appearance of the delta function in \eqref{ql-single} is a
consequence of the assumed uniformity of the magnetic field.  In this
case, $v_\parallel$ is a constant of the unperturbed motion and so a
particle remains in resonance for a long time.  In situations described
by bounce-averaged codes, the magnetic field and $v_\parallel$ vary
along a particle orbit so that the particle does not remain in
resonance.  This effect can be taken into account by averaging
\eqref{ql-single} along a particle trajectory \cite{Bernstein}.  This
removes the delta function, although there are still singularities in
the resulting expression arising from those particles which turn in the
resonance \cite{Kerbel}.

\subsection{Many waves}

\Eqref{ql-single} is easily generalized to include a more realistic
representation of the wave fields.  An important application is to the
incorporation of quasilinear effects into a ray-tracing code.  Here the
externally injected rf power is represented by several rays.  Let us
consider the interaction of these waves with the electrons on a
particular flux surface.  At the point where a given ray intersects the
flux surface it is characterized by its position $\bf r$, wave number
$\bf k$, and power $W$ (usually the frequency $\omega$ is fixed by the
rf source).  $W$ measures the number of watts carried by the ray.  In
order to apply \eqref{ql-single}, we must determine the amplitude of
the corresponding single wave ${\bf E}_w$ which has the same
polarization and same rms field amplitude as the ray (with the rms
averaging performed over time and over the flux surface).

The ray contributes
	$$U={W\over \abs{{\bf v}\!_g\cdot {\bf\hat n}}A_f}$$
	to the wave energy density (in $\rm J/m^3$) averaged over the
flux surface, where ${\bf v}\!_g$ is the group velocity of the ray, $\bf
\hat n$ is the unit vector normal to surface at the point of
intersection and $A_f$ is the area of the flux surface.  The polarization of
the electric field is given by
	$$\mat K\cdot {\bf E}_w=0,$$
	where
	$$\mat K={c^2\over \omega^2}({\bf kk}-k^2\mat I)
+\mat I+i{\bsigma({\bf k},\omega)\over \omega\epsilon_0}$$
	is the dispersion tensor, $c$ is the velocity of light, and
$\bsigma({\bf k},\omega)$ is the conductivity tensor.  The energy
density is related to ${\bf E}_w$ by \cite{Bers}
	$$U={1\over 4}\epsilon_0\omega {\bf E}_w^*\cdot
{\partial \mat K\over \partial \omega}\cdot {\bf E}_w.$$
	Given $W$, we can therefore determine ${\bf E}_w$ (to within an
ignorable phase factor) appropriately averaged over the flux surface.

This is substituted into \eqref{ql-single} and the result summed over
all the rays to give the overall quasilinear diffusion tensor.  In
practice, the delta-functions appearing in this expression must be
replaced by smoothed functions.  This allows the ray-tracing procedure
to reflect the true situation in which a continuous spectrum of waves
is launched.

We complete the discussion of the ray-tracing by pointing out that the
damping of the rays should be calculated self-consistently from $\mat
D_w$.  The power that a particular ray loses per unit volume due to
absorption by the electrons is given by \eqref{pd-def}, where instead
of the total ${\bf S}_w$ we use the contribution the ray in question
makes to ${\bf S}_w$.  To this should be added the power absorbed by the
other species if applicable.  Then the ray power $W$ satisfies the
equation
	$${dW\over dt}=-P\abs{{\bf v}\!_g\cdot {\bf\hat n}}A_f,$$
where the time derivative is the derivative taken along the ray.

If instead of a discrete set of waves, the wave fields are given by a
spectrum
	$${\bf E}({\bf r},t)=
\int {\bf E}_w({\bf k})\exp[i{\bf k\cdot r}-i\omega({\bf k}) t]\,
{d^3{\bf k}\over (2\pi)^3},$$
then \eqref{ql-single} becomes \cite{Kennel}
	$$\mat D_w=\sum_n{q_e^2\over m_e^2}
\int\!{d^3{\bf k}\over (2\pi)^3}{1\over V_p}
\pi\delta[\omega({\bf k})-k_\parallel v_\parallel-n\Omega_e]
{\bf a}_n^\ast{\bf a}_n,
\Eqlab{ql-spectrum}$$
	where $V_p$ is the configuration space volume of the plasma and
the definition of ${\bf a}_n$ is generalized in the obvious way.

\subsection{Model forms}

The results given above allow a ray-tracing code to be coupled to the
solution of the Fokker--Planck equation.  This is an extremely
complicated system, and much work has been carried out using assumed
forms for the quasilinear diffusion coefficient.  This allows us to
study the physics of the interaction of the electrons and the waves
without having to worry about the additional physics of the wave
propagation.  The most widely used model form for lower hybrid waves was
introduced by Fisch \cite{Fisch1} and is given by
	$$\mat D_w=D_w(v_\parallel) {\bf \hat v}_\parallel{\bf \hat v}_\parallel,
\eqn(\eqlab{d-ql}a)$$
where
$$D_w(v_\parallel)=\cases{D_0,&for $v_1<v_\parallel<v_2$,\cr
0,&otherwise.\cr}\eqn(\ref{d-ql}b)$$
	This form of $\mat D_w$ is justified as follows.  Because
lower hybrid waves interact only via the Landau resonance, only the
${\bf \hat v}_\parallel{\bf \hat v}_\parallel$ component is present.  If
$k_\perp v_{te}/\Omega_e\ll 1$, the dependence on perpendicular
velocity may be ignored ($J_0\approx 1$).  Finally, in many cases, the
magnitude of the quasilinear diffusion greatly dominates over the
collisions; thus the quasilinear diffusion coefficient tends to make an
abrupt transition (in velocity space) from being negligible to being
large; if $D_0$ is sufficiently large (i.e., large enough to form a
quasilinear plateau), this situation is accurately modeled by
eq.~(\ref{d-ql}b).

This particular form for $\mat D_w$ is useful because much theoretical
work has been carried out using it \cite{Fisch1}.  Numerical solutions
to the Fokker--Planck equation provide the best test of these
theories. It is therefore important that any numerical code be able to
handle the discontinuities in $\mat D_w$.  (Note, however, that both
$f_e$ and $\bf S$ are continuous even if $\mat D_w$ is not.)

This model is readily generalized, for example, by allowing
$D_0(v_\parallel)$ to be an arbitrary function.  Thus the effect of a
backward component to the lower hybrid spectrum can be studied by
including another boxlike component to $D_0$.  Similar models have
been used to study low-phase-velocity current drive \cite{Fisch-Karney}
and electron-cyclotron current drive \cite{Karney-ec}.

\subsection{Direct specification of the quasilinear flux}

Both analytical and numerical studies show that the current drive
efficiency is primarily determined by the location at which electrons
interact with the waves and the direction in which the waves push the
electrons.  It is sometimes useful to specify the rf-induced flux
directly as some arbitrary vector field ${\bf S}_w({\bf v})$.  Indeed,
in some cases we may know ${\bf S}_w$ more accurately than we know
$\mat D_w$.  In a ray-tracing calculation, $\mat D_w$ may be calculated
self-consistently in terms of the power flows in the various rays.
However, in cruder zero-dimensional calculations, we may wish to assert
merely that so much rf power is absorbed by the electrons.  Then ${\bf
S}_w$ may be estimated from \eqref{pd-def} using an {\sl a priori}
knowledge of which electrons interact with the waves.  Alternatively,
${\bf S}_w$ may be estimated from either an approximate analytic
solution of the Fokker--Planck equation \cite{asymp} or from a solution
of the one-dimensional Fokker-Planck equation \cite{Fisch1}.

If ${\bf S}_w$ is given, then the Fokker--Planck equation (\ref{fp-1}) is
an inhomogeneous (instead of homogeneous) equation.  However, assuming
that one of the linear electron-electron collision operators is being
used, the linear operator acting on $f_e$ in \eqref{fp-1} is now
independent of the wave drive.  This property is used in the adjoint
methods to provide a very efficient method of solving for moments of
$f_e$ (see \secref{adjoint}).

\section{Boundary Conditions}\label{bcs}

\subsection{Computational domain}

We shall take the computational domain $V$ for the Fokker--Planck equation
to be
$$0<v_\perp<v_{\perp\rm max},\qquad
v_{\parallel\rm min}<v_\parallel<v_{\parallel\rm max},
\Eqlab{cyl-domain}$$
	for problems solved in a cylindrical coordinate system and
$$0<v<v_{\rm max},\qquad 0<\theta<\pi,\Eqlab{spher-domain}$$
	for problems solved in a spherical coordinate system.  The
boundary of $V$ is defined to be $A$.  (For example, in spherical
coordinates, $A$ is the spherical surface $v=v_{\rm max}$.)

\subsection{Internal boundaries}

We distinguish two types of boundary:  internal and external boundaries.
The internal boundaries are the simplest.  In a cylindrical
coordinate system we have an internal boundary at $v_\perp=0$.  Values
of $f$ beyond this boundary are determined by symmetry
	$$f_e(-v_\perp,v_\parallel)=f_e(v_\perp,v_\parallel).\Eqlab{cyl-bc}$$
Similarly, in spherical coordinates we have internal boundaries at $v=0$
and at $\theta=0$ and $\theta=\pi$.  These boundaries are treated with
the boundary conditions
	$$\eqalignno{
f_e(-v,\theta)&=f_e(v,\pi-\theta),&(\eqlab{spher-bc}a)\cr
f_e(v,-\theta)&=f_e(v,\theta),&(\ref{spher-bc}b)\cr
f_e(v,\pi+\theta)&=f_e(v,\pi-\theta).&(\ref{spher-bc}c)\cr}$$

\subsection{External boundaries}

The other boundaries are inserted into the problem in violation of the
true physical picture.  In reality the velocity domain extends off to
infinity; on the computer, however, we normally study only a subspace.
We have to choose the subspace to include all the interesting physics:
for studies of electron distribution in a spherical coordinate system,
we require $v_{\rm max}\gg v_{te}$; if the electrons are driven by
lower hybrid waves, then we further require $v_{\rm max}>
(\omega/k_\parallel)_{\rm max}$, the maximum wave phase velocity; if we
wish to study runaways, then $v_{\rm max}$ must exceed the runaway
velocity; and so on.  We next have to choose boundary conditions which
are as ``innocuous'' as possible; i.e., which perturb the solution in
the domain of integration as little as possible compared to the
solution in the full domain.

For electron current-drive problems we choose the condition
	$${\bf S\cdot \hat n}=0,\Eqlab{bc-a}$$
	on the external boundary $A$, where $\bf\hat n$ is the normal
to $A$.  This means that plasma cannot enter or leave the domain of
integration.  Thus the number of electrons is conserved with this
boundary condition.  This boundary condition gives a Maxwellian
steady state in the absence of the rf, and allows a steady-state
solution to be reached in the presence of rf.

If an electric field is present, then in the real problem some
electrons will run away.  Now we wish to impose boundary conditions
which ``allow'' this to happen.  At the boundary we have $v\gg v_{te}$
so that collisions are weak, and the dominant process is the
acceleration by the electric field (we assume that the boundary is
removed from the region where the rf diffusion takes place).  
The Fokker--Planck equation then reduces to a hyperbolic equation.  The
tactic is to apply the same boundary condition as before, namely
\eqref{bc-a}, where the characteristics of the hyperbolic system enter
the domain of integration.  Where the characteristics leave, we set
those diffusion terms which lead to a flux across the boundary to zero.
This makes the equation purely hyperbolic in the direction normal to
the boundary and so {\sl no} boundary condition is required.  (We shall
see in \secref{spatial} how this comes about in the numerical scheme.)

If we assume that $q_eE>0$ so that electrons run away in the positive
direction, then in cylindrical coordinates we would impose
	$$\eqalign{S_\parallel=0,\qquad
&\hbox{for }v_\parallel=v_{\parallel\rm min},\cr
S_\perp=0,\qquad
&\hbox{for }v_\perp=v_{\perp\rm max},\cr
D_{\parallel\perp}=D_{\parallel\parallel}=0,\qquad
&\hbox{for }v_\parallel=v_{\parallel\rm max}.\cr}\Eqlab{bc-e-cyl}$$
(The boundary at $v_\perp=v_{\perp\rm max}$ is taken to be an incoming
boundary because the small collisional friction makes the
characteristics enter along this boundary.)

A slightly more accurate treatment is possible in spherical
coordinates.  If we compare the various collision terms in the
high-velocity limit \eqsref{iso-high}, we find
$F_{cv}\sim D_{c\theta\theta}/v\sim 1/v^2$ and
$D_{cvv}/v\sim v_{te}^2/v^4$.  Thus we can ignore the energy diffusion
term $D_{cvv}$ compared with the other collisional terms.  The
pitch-angle scattering term $D_{c\theta\theta}$ requires no special
handling because it causes diffusion parallel to the boundary.  The
equation is, therefore, hyperbolic in the direction perpendicular to the
boundary with a characteristic acceleration
given by $F_v=F_{cv}+(q_e E/m_e)\cos\theta$.  The boundary conditions
on $v=v_{\rm max}$ then become
	$$\eqalign{S_v=0,\qquad
&\hbox{for }F_{v}<0,\cr
D_{vv}=D_{v\theta}=0,\qquad
&\hbox{for }F_{v}>0.\cr}\Eqlab{bc-e-spher}$$
	For $v_{\rm max}\gg v_{te}$, $F_{cv}$ is accurately
approximated by eq.~(\ref{iso-high}c) (with $a=b=e$).  Thus, if $\abs
E<m_e\Gamma^{e/e}/\abs{q_e}v_{\rm max}^2$, this boundary condition
reduces to \eqref{bc-a}, allowing problems involving both an electric
field and rf diffusion to be handled in a unified way.  In this small
electric field limit, $S_v=0$ is zero everywhere on the boundary and
the numerical runaway rate vanishes.  This is a close approximation to
the true situation in which the runaway rate is exponentially
small---on the order of $\exp(-v_{\rm max}^2/v_{te}^2)$.

\subsection{Treatment of runaways}

With a finite boundary, we can determine the runaway rate accurately
(provided $v_{\rm max}$ is sufficiently large).  However, the behavior
of the runaways beyond the boundary is not followed.  One could, of
course, just choose a very large boundary; but this is wasteful of
computer resources and really just postpones the time at which the
problem is encountered.  It is, therefore, preferable to treat the
runaways as a separate species.  Assuming that the runaways are affected
only by the electric field, the density and current moments of the
runaway population form a closed set of equations.  We define
	$$\eqalign{n_r&=\int_{\overline V} f_e\,\dv,\cr
J_r&=\int_{\overline V} q_ev_\parallel f_e\,\dv,\cr}$$
	where $\overline V$ is the complement of $V$, i.e., the region
$v>v_{\rm max}$ in spherical coordinates.  Applying eqs.~(\ref{cons}a)
and (\ref{cons}b) to $\overline V$ we find
	$$\eqalignno{
{\partial n_r\over \partial t}&=
\int_A{\bf S}\cdot\da,\cr
{\partial J_r\over \partial t}&=
{q_e^2 E\over m_e} n_r+\int_A q_ev_\parallel {\bf S}\cdot\da.\cr}$$
	Thus if we wish to determine the total current as a function of
time, we need only supplement the Fokker--Planck equation by two
ordinary differential equations and then sum the nonrunaway and
runaway contributions to the current.

\section{Spatial Differencing}\label{spatial}
\subsection{Choice of coordinate system}

We have discussed both the cylindrical and the spherical coordinate
systems.  Which one should be used in a given application?  The numerical
scheme that is described here works best if the diffusion tensor is
nearly diagonal.  Then the mixed derivative terms in \eqsref{cylindrical}
or (\ref{spherical}) are small.  (It is these terms which tend to make
the numerical scheme unstable.)  Now the collision operator is
approximately diagonal in spherical coordinates while the quasilinear
term is nearly diagonal in cylindrical coordinates.  Thus the choice of
coordinate system to some extent depends on the relative strength of
these two terms.  Cylindrical coordinates were used in the study of current
drive by low-phase-velocity waves \cite{Fisch-Karney} because the edges
of the resonant region line up with coordinate lines allowing the
scaling with phase velocity to be measured more accurately.  On the
whole, however, the spherical system is to be preferred because the electron-ion
collision term \eqref{coll-ions} becomes large near $v=0$ and we wish this
term to be diagonal.  In \refref{Fisch-Karney} much smaller time
steps had to be taken to avoid the problem with the electron-ion term.
The boundary conditions can also be applied more accurately in
spherical coordinates when an electric field is present
[\eqsref{bc-e-spher}].  For this reason, we will focus on the spherical
coordinate system in this section.  Extension to the cylindrical
coordinate system is straightforward.

An alternate representation of $f_e$ is as a series of Legendre
harmonics.  This has no particular merit in quasilinear problems
because the sharp gradients in $\mat D_w$, \eqref{d-ql}, cause the
Legendre expansion to be slowly convergent.

\subsection{Normalizations}

In solving equations of physical significance on the computer, it is
often useful to normalize all the physical quantities.  This allows us
to work with numbers which are closer to unity (and thus avoid
potential problems due to arithmetic overflow or underflow); more
importantly, the number of parameters needed to specify the problem is
often reduced.

For the problem of current drive by lower hybrid waves, we solve the
Fokker--Planck equation for the electrons.  We normalize velocities to
$v_{te}$ \eqref{vts-max}, times to $\tau_{te}$ \eqref{nute}, the electron
density to $n_e$, the electron distribution to $n_e/v_{te}^3$, the
quasilinear diffusion coefficient to $v_{te}^2\nu_{te}$, the electric
field to $m_e v_{te}\nu_{te}/q_e$, the current density to
$n_eq_ev_{te}$, power density to $n_em_ev_{te}^2\nu_{te}$, etc.

These normalizations coincide with those used by Kulsrud {\sl et al.}\
\cite{Kulsrud}.  However, they differ from those used in some of our
earlier papers, e.g., \refref{Karney-lh}.  (The thermal collision time
differs by a factor of two.)

Since we are only dealing with the electron distribution, we will
drop the species label from $f$ and other electron quantities.
Otherwise, we shall use the same notation for normalized and
unnormalized quantities.  For example, the electron Maxwellian
\eqref{max} reads in normalized terms
	$$f_m(v)={1\over (2\pi)^{3/2}}\exp(-\half v^2).$$

The reduction in the number of parameters now becomes apparent.  The
plasma is characterized by a single parameter $Z_i$ and the quasilinear
diffusion coefficient by three parameters $D_0$, $v_1$, and $v_2$.

\subsection{The numerical grid}

We wish to solve \eqref{fp-1} in the domain $V$ \eqref{spher-domain}.
We do this by converting the differential equation to an algebraic
equation using the finite difference method.  In this method $f$ is
represented by its values on finite set of points and differentials are
represented by differences between neighboring values.

First, we establish a numerical grid by
dividing $v$ and $\theta$ into $N$ and $M$ equal pieces, respectively.
Thus we define
	$$\Delta v=v_{\rm max}/N,\qquad \Delta\theta=\pi/M,\Eqlab{delta-def}$$
	together with grid  positions
	$$\eqalignno{v_j&=j\,\Delta v,&(\eqlab{vj}a)\cr
	\theta_i&=i\,\Delta\theta.&(\ref{vj}b)\cr}$$
	This grid system defines a system of cells.  The electron
distribution function is represented by its values at the {\sl centers}
of these cells, i.e., by the values
	$$f_{i+1/2,j+1/2}=f(v_{j+1/2},\theta_{i+1/2}),
	\quad\hbox{for } 0\le i<M,\quad 0\le j<N,$$
	with $i$ and $j$ being integers; see \figref{grid-fig}.  The
cell $v_j<v<v_{j+1}$, $\theta_i<\theta<\theta_{i+1}$ ($i$ and $j$
integers) is assigned a volume
	$$V_{i+1/2,j+1/2}=2\pi\sin\theta_{i+1/2}v_{j+1/2}^2\,
\Delta v\Delta\theta. \Eqlab{vol}$$
We will define numerical volume integration by
\def\numint{\mathop{\rm int}\nolimits}%
\def\flux{\mathop{\rm flux}\nolimits}%
$$\numint{X}=\sum_{i=0}^{M-1}\sum_{j=0}^{N-1}
X_{i+1/2,j+1/2}f_{i+1/2,j+1/2}V_{i+1/2,j+1/2}.\Eqlab{num-int}$$
	This is the discrete analogue of $\int_V Xf \,\dv$; see
\eqref{int-spher}.  We define the flux of a quantity through
the boundary by
	$$\flux X=\sum_{i=0}^{M-1}
2\pi\sin\theta_{i+1/2} v_N^2 X_{i+1/2,N+1/2}
S_{v,i+1/2,N}\,\Delta\theta,\Eqlab{cons-loss}$$
	which is a discrete analogue of $\int_A X{\bf S}\cdot\da$.  The
number density of electrons becomes
	$$n=\numint1.\Eqlab{n-grid}$$

An alternative approach to finite differences is provided by the
finite-element method where the $f$ is represented by the superposition
of a set of trial functions with finite support.  This approach has been
used in Fokker--Planck codes by workers at Lausanne \cite{Kritz,Succi}.
The finite-element method is also used in some commercial computer
codes for the solution of partial differential equations.  One such code
has been applied to the Fokker--Planck equation by Fuchs {\sl et al.}\
\cite{Fuchs}.  If we identify the weights of the trial functions with
the values of $f$ at the grid positions, we see that the
finite-difference and finite-element methods are quite similar.  In
particular, the goals of the methods are identical: to express
algebraically $\partial f/\partial t$ at a particular location in terms
of $f$ at the same and neighboring locations (usually, the eight
nearest neighbors).  Thus our discussion of the time advancement of the
equation in \secref{time} is independent of the choice of method.

\subsection{Divergence of flux}

Consider the Fokker--Planck equation in the form \eqref{fp-2}.
This is translated onto our numerical grid in a conservative form as
	$$\eqalignno{{\partial f_{i+1/2,j+1/2}\over \partial t}=-\biggl(&
{v_{j+1}^2 S_{v,i+1/2,j+1}-v_j^2 S_{v,i+1/2,j}\over v_{j+1/2}^2 \,\Delta v}\cr
&+
{\sin\theta_{i+1} S_{\theta,i+1,j+1/2}-\sin\theta_i
S_{\theta,i,j+1/2}\over v_{j+1/2}\sin\theta_{i+1/2}\,\Delta \theta}\biggr).
&(\eqlab{flux-grid})}$$
	Notice that the fluxes are required on the edges of the cells
(see \figref{grid-fig}) and that the fluxes on the internal boundaries
do not contribute since they are multiplied by $v_0=0$ or $\sin
\theta_0=\sin\theta_M=0$.  With this method we difference the fluxes
and not the diffusion and friction coefficients.  This lets us treat
problems in which $\mat D_w$ is discontinuous, e.g., as given by
\eqsref{d-ql}.  The scheme in \eqref{flux-grid} is accurate to second
order in $\Delta v$ and $\Delta \theta$.

This form of difference equation is called conservative because it
obeys the conservation law
	$${\partial \numint 1\over \partial t}+\flux 1=0,
\Eqlab{cons-grid}$$
	where $\numint$ and $\flux$ are defined by \eqsref{num-int} and
(\ref{cons-loss}).  This is a discrete counterpart of eq.~(\ref{cons}a).
If $S_{v,i+1/2,N}=0$ for all $i$, then we have $\flux 1=0$ and particles are
exactly conserved in the numerical scheme (if we ignore round-off
errors).  The discrete form of the parallel component of the
momentum conservation law eq.~(\ref{cons}b) is
	$$\eqalignno{{\partial \numint (v\cos\theta)\over \partial t}
+\flux (v\cos\theta)
&=\sum_{i=0}^{M-1}\sum_{j=0}^N
2\pi v_j^2\sin\theta_{i+1/2}\cos\theta_{i+1/2} S_{v,i+1/2,j}
\,\Delta v\Delta\theta\qquad\cr
&\quad{}-\sum_{i=0}^{M}\sum_{j=0}^{N-1}
2\pi v_{j+1/2}^2\sin^2\!\theta_{i}S_{\theta,i,j+1/2}
\,\Delta v\,2\sin(\half\Delta\theta),&(\eqlab{momcons-grid})\cr}$$
while the energy conservation relation eq.~(\ref{cons}c) becomes
	$${\partial \numint(\half v^2)\over\partial t}
+\flux(\half v^2)=
\sum_{i=0}^{M-1}\sum_{j=0}^{N}
2\pi\sin\theta_{i+1/2} v_j^3 S_{v,i+1/2,j} \,\Delta v\Delta\theta.
\Eqlab{encons-grid}$$
	These relations are useful in that they establish definitions
of various physical quantities that are consistent with the numerical
scheme.  For example, we can interpret the right-hand side of
\eqref{encons-grid} as the total power flowing into the electrons.  This
definition is consistent with the numerical definition of the energy of
the electrons, namely $\numint(\half v^2)$.  Furthermore, we can
determine the power flowing into the electrons from the waves (for
example) by replacing $S_v$ in the right-hand side of this equation by
the flux due to the waves $S_{wv}$ [compare with
\eqref{pd-def}].  In this way, we obtain a complete and accurate power
balance for the electrons.  Similarly, the right-hand side of
\eqref{momcons-grid} gives the definition of the force on the
electrons.  This is used when evaluating $P_\parallel^{e/e}$ in
\eqref{force-def}.

[In order to prove \eqsref{momcons-grid} and (\ref{encons-grid}), the
following relation is useful:
	$$\sum_{i=0}^{M-1}
\half(A_{i+1}+A_i)(B_{i+1}-B_i)
=A_MB_M-A_0B_0-\sum_{i=0}^{M-1}
\half(B_{i+1}+B_i)(A_{i+1}-A_i).$$
This is the rule for  ``summing by parts''---the discrete counterpart
of integration by parts.]

The basic difference equation (\ref{flux-grid}) is readily generalized
to nonuniform grids.  However, the derivation of \eqsref{momcons-grid}
and (\ref{encons-grid}) relies on the uniformity of the grid and they
cannot easily be generalized.  Nonuniform spacing is used in
{\sc FPPAC} \cite{McCoy}.

\subsection{Stream function}

A very useful tool for understanding the Fokker--Planck equation
(\ref{fp-2}) is the flux plot, which shows the vector field ${\bf
S}({\bf v})$.  This is sometimes displayed as a set of arrows, one at
each grid point, which point in the direction of ${\bf S}$ and which
have a length proportional to $S$.  In this problem, $S_v$ and
$S_\theta$ are known at different locations, so that realization of
this prescription would necessitate interpolation.  Furthermore, such a
display is often very misleading because the visual impression is
strongly affected by whether the arrows line up with other grid points
or not---a purely artificial aspect of the problem.

The much superior method is possible if we restrict ourselves to the
steady state.  In this case, the vector field ${\bf S}({\bf v})$ is
divergence-free $\nabla\cdot{\bf S}=0$, and so may be expressed as the
curl of a stream function, i.e.,
	$${\bf S}({\bf v})=\nabla\times{nA({\bf v}){\bf \hat{\bphi}}
\over 2\pi v\sin\theta},$$
	where $\phi$ is the azimuthal coordinate.  The components of
$\bf S$ are given by
	$$\eqalignno{
S_v&={n\over 2\pi v^2\sin\theta}{\partial A\over\partial \theta},
&(\eqlab{stream}a)\cr
S_\theta&=-{n\over 2\pi v\sin\theta}{\partial A\over\partial v}.
&(\ref{stream}b)\cr}$$
	Because ${\bf S}\cdot\nabla A=0$, lines of constant $A$ are
stream lines.  Thus a contour plot of $A({\bf v})$ gives the vector
field of ${\bf S}({\bf v})$.  The stream lines are obviously closed
(indicating that the flow is divergence-free), and the total flux of
electrons between any two contours is equal to the difference in the
values of $nA$ on those two contours.

We can compute $A$ on the numerical grid using discrete analogs of
\eqsref{stream}
	$$\eqalignno{A_{i,j}&={2\pi v_j^2\over n}\sum_{i'=0}^{i-1}
\sin\theta_{i'+1/2}S_{v,i'+1/2,j}\,\Delta\theta,
&(\eqlab{stream-grid}a)\cr
&=-{2\pi\sin\theta_i\over n}\sum_{j'=0}^{j-1}
v_{j'+1/2}S_{\theta,i,j'+1/2}\,\Delta v.
&(\ref{stream-grid}b)\cr}$$
	If $\partial f_{i+1/2,j+1/2}/\partial t=0$ according to
\eqref{flux-grid}, then these two definitions are consistent.

\subsection{Computation of the flux}

In order to complete the specification of the difference scheme we must
give formulas for $S_{v,i+1/2,j}$ and $S_{\theta,i,j+1/2}$ in
\eqref{flux-grid}.  These depend on the type of electron-electron
collision operator used.  We start with collisions off a
Maxwellian background $C_{\rm Max}^{e/e}$, \eqref{coll-max}.  This is
the simplest case and yet it exhibits all the difficulties of solving
the Fokker--Planck equation.

The collisional flux is given by the sum of the flux contributing to
$C_{\rm Max}^{e/e}$ which is given by \eqsref{isotropic} and
(\ref{iso-max}) and the flux contributing to $C^{e/i}$ which is given
by \eqsref{isotropic} and (\ref{ions}).  [In fact, we compute the
electron-electron flux by numerically evaluating the integrals in
\eqsref{iso-gen}.]  To this is added
the quasilinear flux from \eqsref{wave-flux} and (\ref{d-ql}) and the
electric-field-induced flux from \eqref{s-e}.  Both these terms are
converted into spherical coordinates using \eqsref{transform}.  The
total flux is then given by the general equations (\ref{spherical}b)
and (\ref{spherical}c).

The diffusion and friction coefficients are computed at the points at
which we need to know $S_v$ and $S_\theta$.  Thus we compute
$D_{vv,i+1/2,j}$, $D_{v\theta,i+1/2,j}$, $F_{v,i+1/2,j}$, and
$D_{\theta v,i,j+1/2}$, $D_{\theta\theta,i,j+1/2}$,
$F_{\theta,i,j+1/2}$.
The coefficients for $S_v$ are not required at $j=0$, nor
those for $S_\theta$ at $i=0$, $M$, because these fluxes are multiplied
by zero in \eqref{flux-grid}.  The boundary conditions \eqsref{bc-e-spher}
at $v_{\rm max}$ are handled by setting
	$$\eqalign{
D_{vv,i+1/2,N}&\leftarrow0,\cr
D_{v\theta,i+1/2,N}&\leftarrow0,\cr
F_{v,i+1/2,N}&\leftarrow\max(F_{v,i+1/2,N},0).\cr}$$

Next we must specify the way in which $f$ and its derivatives are to be
computed at the edges of the cells---i.e., locations $(i+1/2,j)$ and
$(i,j+1/2)$---in terms of the values of $f$ at the centers of the cells
$(i+1/2,j+1/2)$.  Two of the terms are straightforward:
	$$\eqalignno{
{\partial f_{i+1/2,j}\over\partial v}&
={f_{i+1/2,j+1/2}-f_{i+1/2,j-1/2}\over \Delta v},&(\eqlab{deriv}a)\cr
{\partial f_{i,j+1/2}\over\partial \theta}&
={f_{i+1/2,j+1/2}-f_{i-1/2,j+1/2}\over \Delta
\theta}.&(\ref{deriv}b)\cr}$$
	Again these expressions are accurate to second order.

The evaluation of $f$ at the cell edges uses a method proposed by
Chang and Cooper \cite{Chang} extended here to
two dimensions.  The simple method, i.e., 
$$f_{i+1/2,j}=
\half(f_{i+1/2,j+1/2}+f_{i+1/2,j-1/2}),$$  turns out to give poor
results for the steady-state distribution.  Chang and Cooper replace
this with
	$$\eqalignno{
f_{i+1/2,j}&=(1-\delta_{i+1/2,j}) f_{i+1/2,j+1/2}
+\delta_{i+1/2,j} f_{i+1/2,j-1/2},&(\eqlab{value}a)\cr
f_{i,j+1/2}&=(1-\delta_{i,j+1/2}) f_{i+1/2,j+1/2}
+\delta_{i,j+1/2} f_{i-1/2,j+1/2},&(\ref{value}b)\cr}$$
	where the $\delta$s are given by
	$$\eqalignno{
\delta_{i+1/2,j}&=g(-\Delta v F_{v,i+1/2,j}/D_{vv,i+1/2,j}),
&(\eqlab{del-def}a)\cr
\delta_{i,j+1/2}&=g(-\Delta \theta F_{\theta,i,j+1/2}/D_{\theta\theta,i,j+1/2}),
&(\ref{del-def}b)\cr}$$
and
	$$g(w)={1\over w}-{1\over \exp(w)-1}.\Eqlab{g-def}$$

The role of the $\delta$ is to weight the averaging performed in
\eqsref{value}.  The weighting is needed because often $f$ is a strongly
(exponentially) varying function of ${\bf v}$.  An acute example of this
is the Maxwellian distribution which varies very strongly for large
$v$.  In fact, the weighting is such that a Maxwellian is an {\sl exact}
steady-state solution when there is no rf and no electric field and
when $C_{\rm Max}^{e/e}$ is employed as the electron-electron collision
operator.  This is easily seen because for any isotropic distribution
$S_{c\theta}=0$; in that case, we also require $S_{cv}=0$ in the steady
state (because there are no sources or sinks of electrons).  Using
eqs.~(\ref{isotropic}a) (with $a=b=e$), (\ref{deriv}a), and
(\ref{value}a), together with
$F_{cv,i+1/2,j}^{e/e}/D_{cvv,i+1/2,j}^{e/e}=-v_j$, we find
	$${f_{i+1/2,j+1/2}\over f_{i+1/2,j-1/2}}
={f_{m,j+1/2}\over f_{m,j-1/2}}
=\exp(-v_j\,\Delta v).$$
	The errors in various moments of $f$ are, therefore, exponentially
small.  With one-dimensional equations the weighting cures the problem
of $f$ becoming negative \cite{Chang}.  With our two-dimensional
equation, this problem is alleviated but not cured.  In general, this
problem is solved by taking a sufficiently fine mesh (assuming that the
electron-electron collision operator preserves the non-negative nature
of $f$).

The function $g$ has the properties
	$$\eqalign{g(w)&=1-g(-w),\cr
g(w)&={1\over 2}-{w\over 12}+{w^3\over 720}+\ldots,\cr
g(-\infty)&=1,\qquad
g(0)={1\over 2},\qquad g(\infty)=0.\cr}$$
	The first two properties are useful for evaluating $g(w)$ for
$w\gg1$ and $w\approx 0$, respectively.

The values of the cross-derivative terms which multiply the
off-diagonal terms in the diffusion tensor ($D_{v\theta}$ and
$D_{\theta v}$) are now given in terms of \eqsref{value} as
	$$\eqalignno{
{\partial f_{i+1/2,j}\over \partial \theta}&=
{f_{i+3/2,j}-f_{i-1/2,j}\over 2\Delta\theta},&(\eqlab{cross}a)\cr
{\partial f_{i,j+1/2}\over \partial v}&=
{f_{i,j+3/2}-f_{i,j-1/2}\over 2\Delta v}.&(\ref{cross}b)\cr}$$

The internal boundary conditions \eqsref{spher-bc} give the values of
$f_{i+1/2,j+1/2}$ beyond the internal boundaries as
	$$\eqalign{f_{i+1/2,-1/2}&=f_{M-i-1/2,1/2},\cr
f_{-1/2,j+1/2}&=f_{1/2,j+1/2},\cr
f_{M+1/2,j+1/2}&=f_{M-1/2,j+1/2}.\cr}$$
	These conditions are only needed in the evaluation of
cross-derivative terms.  The form of \eqref{flux-grid} automatically
takes care of the internal boundaries for the other terms.

The external boundary at $v=v_{\rm max}$ is treated as follows: In the
computation of $S_{v,i+1/2,N}$ we need only worry about the friction
term (since $D_{vv}=D_{v\theta}=0$ on the boundary) so that only
$f_{i+1/2,N}$ is needed.  Furthermore, the friction coefficient
$F_{v,i+1/2,N}$ is non-negative.  From eq.~(\ref{value}a), we have
$f_{i+1/2,N}=f_{i+1/2,N-1/2}$ because $\delta_{i+1/2,N}\rightarrow1$
for $F_{v,i+1/2,N}>0$ and $D_{vv,i+1/2,N}=0+$.  (Obviously the value of
$f_{i+1/2,N}$ is not required where $F_{v,i+1/2,N}=0$.)  Recall that the
equation reduces to hyperbolic type on this boundary, so that no
boundary condition should need to be specified here, as indeed is the
case.  In fact, the method reduces to the standard upstream differencing
for a hyperbolic equation on this boundary.  In the computation of
$S_{\theta,i,N-1/2}$, only the cross-derivative term $\partial
f_{i,N-1/2}/ \partial v$ potentially involves points outside the
integration domain.  In this term, we use
	$${\partial f_{i,N-1/2}\over \partial v}=
{f_{i,N-1/2}-f_{i,N-3/2}\over \Delta v},$$
	instead of eq.~(\ref{cross}b).

\subsection{Matrix formulation}

For collisions off a Maxwellian background the problem is {\sl linear}
so that \eqref{flux-grid} can be rewritten as
$${\partial f\over \partial t}+ Af=h,\Eqlab{matrix-eq}$$
	where $f$ is a vector of length $MN$ of the values
$f_{i+1/2,j+1/2}$ and $A$ is an $MN\times MN$ matrix of coefficients.
The right-hand side $h$ (also a vector of length $MN$) is inserted to
aid in the treatment of other collision operators.  For the Maxwellian
collision operator, we have $h=0$.  It is convenient to split $A$ into
three pieces, namely
	$$A=A_v +A_\theta +A_\times,$$
	where $A_v$ contains the terms proportional to $D_{vv}$ and
$F_v$, $A_\theta$ contains those proportional to $D_{\theta\theta}$
and $F_\theta$, and $A_\times$ contains the cross-derivative terms
proportional to $D_{v\theta}$ and $D_{\theta v}$.  With the difference
scheme given in this section $A_v$ and $A_\theta$ are tridiagonal
matrices.  Thus we can write
	$$\eqalignno{(A_vf)_{i+1/2,j+1/2}
&= a_{v,i+1/2,j+1/2}f_{i+1/2,j-1/2}
 +b_{v,i+1/2,j+1/2}f_{i+1/2,j+1/2}\cr
&\quad +c_{v,i+1/2,j+1/2}f_{i+1/2,j+3/2},&(\eqlab{tridiag}a)\cr
(A_\theta f)_{i+1/2,j+1/2}
&= a_{\theta,i+1/2,j+1/2}f_{i-1/2,j+1/2}
 +b_{\theta,i+1/2,j+1/2}f_{i+1/2,j+1/2}\cr
&\quad +c_{\theta,i+1/2,j+1/2}f_{i+3/2,j+1/2},&(\ref{tridiag}b)\cr}$$
where
$$\eqalignno{
a_{v,i+1/2,j+1/2}&=
{v_j^2\over B_v}
\biggl(-{D_{vv,i+1/2,j}\over\Delta v}
-F_{v,i+1/2,j}\delta_{i+1/2,j}\biggr),&(\eqlab{mat-coeff}a)\cr
b_{v,i+1/2,j+1/2}&=
{v_j^2\over B_v}
\biggl({D_{vv,i+1/2,j}\over\Delta v}
-F_{v,i+1/2,j}\epsilon_{i+1/2,j}\biggr)\cr
&\quad{}+{v_{j+1}^2\over B_v}
\biggl({D_{vv,i+1/2,j+1}\over\Delta v}
+F_{v,i+1/2,j+1}\delta_{i+1/2,j+1}\biggr),&(\ref{mat-coeff}b)\cr
c_{v,i+1/2,j+1/2}&=
{v_{j+1}^2\over B_v}
\biggl(-{D_{vv,i+1/2,j+1}\over\Delta v}
+F_{v,i+1/2,j+1}\epsilon_{i+1/2,j+1}\biggr),&(\ref{mat-coeff}c)\cr
a_{\theta,i+1/2,j+1/2}&=
{\sin\theta_i\over B_\theta}
\biggl(-{D_{\theta\theta,i,j+1/2}\over v_{j+1/2}\Delta \theta}
-F_{\theta,i,j+1/2}\delta_{i,j+1/2}\biggr),&(\ref{mat-coeff}d)\cr
b_{\theta,i+1/2,j+1/2}&=
{\sin\theta_i\over B_\theta}
\biggl({D_{\theta\theta,i,j+1/2}\over v_{j+1/2}\Delta \theta}
-F_{\theta,i,j+1/2}\epsilon_{i,j+1/2}\biggr)\cr
&\quad{}+{\sin\theta_{i+1}\over B_\theta}
\biggl({D_{\theta\theta,i+1,j+1/2}\over v_{j+1/2}\Delta \theta}
+F_{\theta,i+1,j+1/2}\delta_{i,j+1/2}\biggr),&(\ref{mat-coeff}e)\cr
c_{\theta,i+1/2,j+1/2}&=
{\sin\theta_{i+1}\over B_\theta}
\biggl(-{D_{\theta\theta,i+1,j+1/2}\over v_{j+1/2}\Delta \theta}
+F_{\theta,i+1,j+1/2}\epsilon_{i,j+1/2}\biggr),&(\ref{mat-coeff}f)\cr}$$
	where $\epsilon=1-\delta$, $B_v=\Delta v\,v_{j+1/2}^2$, and
$B_\theta=v_{j+1/2}\Delta \theta\,\sin\theta_{i+1/2}$.  With these
coefficients the boundary conditions are reflected in the relations
$a_{v,i+1/2,1/2}=c_{v,i+1/2,N-1/2}=0$ and
$a_{\theta,1/2,j+1/2}=c_{\theta,N-1/2,j+1/2}=0$, which are automatically
satisfied.

The matrix $A_\times$ is more complicated with $(A_\times
f)_{i+1/2,j+1/2}$ depending, in general, on the eight nearest neighbors
to $f_{i+1/2,j+1/2}$.  The boundary conditions have to be explicitly
included in this matrix.  We do not give expressions for the components
of $A_\times$ here because only the product $A_\times f$ is ever needed
in the calculation.  This is most easily computed directly in terms
of the flux; this also cuts down on the storage requirements.

\subsection{Alternate collision operators}

The methods we will describe in the next sections for solving
\eqref{matrix-eq} depend on the linearity of this equation and 
the fact that $A_v$ and $A_\theta$ are tridiagonal matrices.  With
more complicated electron-electron collision operators, these
conditions no longer hold.  However, the techniques can still be used
because the difference between the other collision terms and the
Maxwellian collision term varies slowly in time.

If the full electron-electron collision operator is used, the basic
framework given above still applies, except that the diffusion and
friction coefficients $\mat D_c^{e/e}$ and ${\bf F}_c^{e/e}$ are now
given in terms of gradients of the Rosenbluth potentials
\eqsref{coll-ros}.  These coefficients depend on $f$ making the equation
nonlinear.  In practice, the dependence on $f$ is weak so that the
coefficients only need to be recomputed occasionally.  This also means
that the equation is approximately linear so that the linear matrix
techniques used to advance the equation in time still apply.

If the linearized or truncated collision operators are used, then the
equation remains linear but with a term which involves an integral over
$f$, namely $C\bigl(f_m(v),f({\bf v})\bigr)$ or the truncation of this term.
Again, this term is weakly dependent on $f$ so that it need not be
recomputed every time step.  It is then most convenient to regard this
term as the inhomogeneous driving term $h$ \eqref{matrix-eq}.  For
the truncated collision operator $C_{\rm trunc}^{e/e}$,
\eqref{truncated}, the elements of $h$ are given
by $C\bigl(f_m(v),f^{(1)}(v)\cos\theta\bigr)$ evaluated at
$(v_{j+1/2},\theta_{i+1/2})$.  The computation of this term is described
in \appref{numerical}.

\section{Time Differencing}\label{time}

\subsection{Crank--Nicholson method}

We now turn to the method for advancing the Fokker--Planck equation in
time.    If the time step is $\Delta t$, then we define
	$$f^k=f(t=t_k),\qquad t_k=k\Delta t.\Eqlab{time-step}$$
	The simplest way of advancing \eqref{matrix-eq} is the explicit
scheme
	$${f^{k+1}-f^k\over \Delta t}+Af^k=h.$$
	This is only accurate to first order in $\Delta t$.  Furthermore,
$\Delta t$ must be chosen to be very small, on the order of $\Delta v^2$ or
$\Delta \theta^2$, for stability.  These defects are easily remedied by
the Crank--Nicholson scheme \cite{Marchuk} which reads
	$${f^{k+1}-f^k\over \Delta t}+A{f^{k+1}+f^k\over 2}=h.
\Eqlab{crank}$$
	This scheme is accurate to second order in $\Delta t$ and
is stable if $A$ is positive definite.  (This is a
condition possessed by the continuous form of the operator $A$.)  In
order to solve \eqref{crank} for $f^{k+1}$ we have to compute the
inverse of $(I+\half\Delta t\,A)$.  This is a large banded matrix which
can either be inverted using iterative methods or using Gaussian
elimination.  In both cases the number of operations is $O(N^3)$,
(assuming $M\sim N$) making it a very expensive proposition.
(This approach is discussed further in \secref{steady-state}.)

\subsection{Alternating-direction-implicit method}

Although $(I+\half\Delta t\,A)$ is difficult to invert, the matrices
$(I+\half\Delta t\,A_v)$ and $(I+\half\Delta t\,A_\theta)$ are rather
easily inverted.  This allows the {\sl alternating-direction-implicit}
method \cite{Marchuk} to be used.  Unfortunately, $(I+\half\Delta
t\,A_\times)$ is not easily inverted and this means that the
cross-derivative terms are treated explicitly in this method.  Consider
the equation
	$$\biggl(I+{\Delta t\over2}A_v\biggr)
\biggl(I+{\Delta t\over2}A_\theta\biggr){f^{k+1}-f^k\over\Delta t}
+Af^k=h.\Eqlab{split}$$
	If we rearrange the terms in this equation to give
	$$\biggl(I+{\Delta t^2\over 4}A_vA_\theta\biggr)
{f^{k+1}-f^k\over\Delta t}
+(A_v+A_\theta){f^{k+1}+f^k\over 2}+A_\times f^k=h,$$
	we see that this method differs from the Crank--Nicholson
method in two respects.  Firstly, there is a $\Delta t^2$ term
multiplying the time difference term.  This difference is unimportant
because it does not alter the accuracy of the scheme.  Secondly, the
cross-derivative terms are treated explicitly.  If we ignore the
cross-derivative terms, \eqref{split} is as accurate as the
Crank--Nicholson scheme, but is much easier to realize because it is
easy to solve \eqref{split} for $f^{k+1}$.  The explicit treatment of the
cross-derivative terms lowers the accuracy and the stability, putting a
limit on the maximum $\Delta t$ that can be used.  On the other hand,
the implicit treatment of the other terms means that this method is far
superior to the fully explicit method.

We can compute $f^{k+1}$ from \eqref{split} in a series of simple steps:
$$\eqalign{\phi^k&=h-Af^k,\cr
\xi^{k+1/2}&=\biggl(I+{\Delta t\over2}A_v\biggr)^{-1}\phi^k,\cr
\xi^{k+1}&=\biggl(I+{\Delta t\over2}A_\theta\biggr)^{-1}\xi^{k+1/2},\cr
f^{k+1}&=f^k+\Delta t\,\xi^{k+1}.\cr}$$
	The inversion of the matrices is carried out using Gaussian
elimination as described in \appref{numerical}.

\subsection{Example}

Let us consider a specific example relevant to lower hybrid current
drive.  The plasma consists of electrons and infinitely massive ions
with $Z_i=1$.  Electron-electron collisions are computed assuming a
Maxwellian background using $C_{\rm Max}^{e/e}$ \eqref{coll-max}.
Electron-ion collisions are given by \eqref{coll-ions}.  The effect of
the lower hybrid waves is modeled by a quasilinear diffusion
coefficient given by \eqsref{d-ql} with $D_0=1$, $v_1=3$, and $v_2=5$.
The electric field $E$ is taken to be zero.  Except for minor details
this is the same example treated in the paper on lower hybrid
current drive \cite{Karney-lh}.  (The time normalization used
in that paper differs from the one adopted here by a factor of two.)  We
take $f(t=0)=f_m$, $v_{\rm max}=10$, $M=N=100$, and $\Delta t=0.2$.

In studies of current drive, we are principally interested in the
current density $J$, the rf power absorbed per unit volume by the
plasma $P$, and their ratio $J/P$.  These are defined by
\eqsref{curr-def} and (\ref{pd-def}) whose discrete forms read
	$$\eqalignno{
J&={\numint(v\cos\theta)\over n},&(\eqlab{curr-def-grid})\cr
P&={1\over n}\sum_{i=0}^{M-1}\sum_{j=0}^{N}
2\pi\sin\theta_{i+1/2} v_j^3 S_{wv,i+1/2,j} \,\Delta v\Delta\theta,
&(\eqlab{pd-def-grid})\cr}$$
	    where $n$ is given by \eqref{n-grid}.  (These definitions
include a $1/n$ factor, because the $n$ is included in the
normalizations for $J$ and $P$.)

The current is plotted as a function of time in
\figref{current}.  With $\Delta t=0.5$, the integration is unstable.  The
difference in the values of the current when the equations are
integrated with $\Delta t=0.2$ and $\Delta t=0.05$ is about $0.1\%$ of
the final current.

The steady-state solution for $f$ is shown in \figref{f-steady}.  This
may be obtained by integrating the equation sufficiently long (until
about $t=1000$) with a fixed time step or else using the techniques
described in \secref{steady-state}.  (With this numerical method, the
steady state is independent of $\Delta t$.)  The plateau in the resonant
region is clearly visible as well as the considerable perpendicular
heating.  Using \eqsref{curr-def-grid} and (\ref{pd-def-grid}), we have
$J=5.754\times10^{-2}$, $P=4.011\times10^{-3}$, and $J/P=14.34$.

The flux plot for this case is given in \figref{flux-fig}.  This shows
that the combination of rf diffusion and collisional scattering induces
a perpendicular flux in the resonant region.  Such flux plots are
useful in providing guidance for the analytic solution of this problem
\cite{asymp}.
More extensive examination of this example can be found in the original
paper \cite{Karney-lh} including projections onto the $v_\parallel$
axis, slices at constant $v_\perp$, etc.

There are two possible sources of error in these results: errors
arising from the finite boundary (i.e., because $v_{\rm max}$ is
finite) and errors arising from the finite mesh.  The effect of the
boundary can be determined by increasing $v_{\rm max}$ to 20 (and
increasing $N$ to 200).  In the steady state, this gives
$J=5.759\times10^{-2}$, $P=4.012\times10^{-3}$, $J/P=14.35$---changes
of less than $0.1\%$.  Thus for this particular problem, $v_{\rm
max}=10$ is adequate.

The effect of the discrete spatial grid is found by varying $\Delta v$
and $\Delta \theta$.  This we do by keeping $v_{\rm max}=10$ varying $M$
and $N$ with $M=N$.  Thus we have $N=10/\Delta v$ and
$\Delta\theta=\Delta v \,\pi/10$.  The results for $J$ and $J/P$ are
shown in \figref{conv-d}.  We see that there is a lot of scatter in the
data which arises because $\mat D_w$ is discontinuous.  As $\Delta v$
and $\Delta \theta$ are varied, grid points (those on which the flux is
defined) enter or leave the resonant region $v_1<v_\parallel<v_2$.  Each
time this happens, there is a jump in $J$ and $P$.  As $\Delta
v\rightarrow 0$, $J$ approaches its asymptotic value of about
$5.6\times10^{-2}$ and the convergence to this value is as $\Delta v$.
The finite mesh error in $J$ with $M=N=100$ is about $3\%$.  This rate
of convergence can be understood because $J$ and $P$ are exponentially
dependent on $v_1$ [$J\sim \exp(-\half v_1^2)$] and $v_1$ is determined
only to within $\pm\half\Delta v$.  Thus the relative error in $J$ and
$P$ is about $\exp(\half v_1\Delta v)-1\approx \half v_1\Delta v$.  This
gives a relative error of $15\%$ for $v_1=3$, $N=100$, $v_{\rm
max}=10$.  The actual error is somewhat less than this because the
boundary of the resonant region cuts across the grid lines and so $v_1$
is in fact determined more accurately than was assumed here.  Because
$J$ and $P$ are both subject to the same error, the ratio $J/P$ is more
accurately given: convergence to the asymptotic value of $14.24$ is as
$\Delta v^2$ and the value with $M=N=100$ is in error by less than $1\%$.

If instead we use the truncated electron-electron collision operator $C_{\rm
trunc}^{e/e}$, the steady-state distribution function is
rather similar to that shown in \figref{f-steady}.  However, the flux
plot \figref{trunc-flux} shows a new eddy at low velocities due to the
overall drift of the electrons with respect to the ions.  (This plot is
obtained with the same parameters as for \figref{flux-fig}.)  In this
case, we find $J=7.092\times10^{-2}$, $P=4.294\times10^{-3}$,
$J/P=16.52$.  The enhancement of the efficiency $J/P$ comes about
because momentum (and hence current) is no longer lost when tail
electrons collide with bulk electrons.

A check on the implementation of the $C_{\rm trunc}^{e/e}$ is given by
measuring the electrical conductivity.  For $Z_i=1$, the exact
conductivity is given by \tabref{conduct} as $J/E=7.429\approx
0.582\times16\sqrt{2/\pi}$ \cite{Spitzer}.  Integrating the
Fokker--Planck equation using the truncated collision operator with no
rf $D_0=0$ and a small electric field $E=10^{-3}$, the conductivity is
$J/E=7.446$, a $0.3\%$ error.  This small error is probably attributable
partly to the finite mesh size (here we again took $M=N=100$ and
$v_{\rm max}=10$) and partly to the finiteness of $E$ (since there is a
contribution to the current which varies as $E^3$).  In contrast, if
$C_{\rm Max}^{e/e}$, is used the conductivity is $J/E=3.772$ a factor of
two too small \cite{Kulsrud}.

\section{Steady-State Solution}\label{steady-state}

\subsection{Statement of problem}

Often, we are only interested in the steady-state solution to
the Fokker--Planck equation.  Nearly always we must resort to an
iterative method for obtaining the steady state.  In that case we need
some measure of how close we are to the steady state so that iteration
may be stopped when this is small enough.  The measure we shall employ is
	$$R={1\over n}
\sqrt{\numint\biggl[\biggl({\partial f\over\partial t}\biggr)^2\biggr]},
\Eqlab{residue}$$
	where the {\sl residue} $\partial f/\partial t$ is given by
\eqref{flux-grid}.  Somewhat arbitrarily we use $R=10^{-9}$ as the
convergence criterion.

One obvious way of obtaining a steady state is to integrate the
time-depen\-dent solution as described in \secref{time} for a long time.
This should be done with the largest time step consistent with
stability.  For the example shown in \figref{f-steady}, the
convergence criterion is met at time $t=812$.  The largest time step
that can be used is approximately $0.2$; so that 4060 steps are
required.  The {\sc CPU} time required to run the Fokker--Planck code on
the Cray--1 is approximately $2\,\mu \rm s$ per mesh point per time
step.  Thus, achieving the steady state by this method takes about
$80\,\rm s$.  This is rather expensive and it is therefore desirable to
find faster methods.

However, this method is quite effective when
$A_\times=0$.  Then the numerical scheme is stable even if $\Delta t$ is
large.  For example, for the electric field example discussed in
\secref{time} in which $D_0=0$ and $E=10^{-3}$, we can take $\Delta t
= 1$, and the convergence criterion is met after 220 steps.  Here
the integral portion of $C_{\rm trunc}^{e/e}$, which is represented by
the term $h$ in \eqref{matrix-eq}, is evaluated every tenth time step.
The numerical method is stable for larger values of $\Delta t$.  But,
because the integration is less accurate, more steps are required to
meet the convergence criterion.  With large $\Delta t$ the numerical
solution tends to oscillate about the steady state.

\subsection{Chebyshev acceleration}

A significant improvement can be achieved by using a varying time step.
Hewett {\sl et al.}\ \cite{Hewett} describe an adaptive time selection
for the alternating direction implicit method which speeds the
convergence by a factor of two to three.  Here we describe Chebyshev
acceleration \cite{Marchuk} which is a nonadaptive method for
selecting varying time steps.  We choose the time step
$\Delta t_k=t_{k+1}-t_k$ according to 
	$$\Delta t_k={2\over
\displaystyle
\beta+\alpha-(\beta-\alpha)\cos\biggl({[2(k\bmod K)+1]\pi
\over 2K}\biggr)},
\Eqlab{Tcheb-dt}$$
	where $\alpha$, $\beta$, and $K$ are constants with
$\alpha<\beta$ and $K={\rm integer}$.  The advantage of this method is
that by changing a few lines of code it can easily be incorporated into
the alternating-direction-implicit method described in \secref{time}.
A fixed time step is recovered in the special case
$\alpha=\beta=1/\Delta t$.

Let us discuss the choice of the parameters in \eqref{Tcheb-dt}.  With
$K$ large, \eqref{Tcheb-dt} gives a series of $K$ time steps (repeated
periodically) varying from $1/\alpha$ down to $1/\beta$.  In the
examples we consider, we take $K=20$.  Then the maximum time
step is somewhat less than $1/\alpha$ while the minimum time step is
very close to $1/\beta$.  In order to realize performance gains with
this method we wish to pick the minimum time step comfortably within
the stability threshold for the fixed-time-step method, while the
maximum time step is considerably greater than the stability
threshold.

The method works because the long wavelength eigenmodes of the linear
operator decay slowly but are stable with large $\Delta t$; on the
other hand, the
short wavelength modes decay rapidly but are only stable if $\Delta t$
is small.  Consider a particular cycle of $K$ steps.  During the initial
large time steps, the long wavelength modes are efficiently damped
(because $\Delta t$ is large), but the short wavelength modes grow.
This is followed by successively shorter time steps which damp the
short wavelength modes.

For the example shown in \figref{f-steady}, the stability threshold for
$\Delta t$ lies between 0.2 and 0.5.  Thus we choose $1/\beta=0.05$ and
$1/\alpha=1000$.  With $K=20$ this gives a maximum time step of 31.4, a
minimum step of 0.05, and an average time step of 1.95.  Since the
average time step is about 10 times the largest time step that can be
used in the fixed time step scheme, we expect convergence to be 10
times faster.  Indeed this is the case.  The convergence criterion is met
after 400 steps at $t=790$.  This takes about $8\,\rm s$ of {\sc CPU}
time.  The variation of $R$ with time is shown in \figref{residu-fig}.
This shows the growth of $R$ during the large time steps followed by a
drop in $R$ as the instabilities are quenched during the small time
steps.  The overall decay of $R$ with $t$ closely matches that seen with
a fixed time step.  (This is contrary to the experience of Hewett {\sl
et al.}\ with their adaptive code in which the rates of decay are very
different \cite{Hewett}.)

\subsection{Runaway problem}

If the electric field is sufficiently large to produce runaways, i.e.,
$E>v_{\rm max}^{-2}$, then as $t\rightarrow\infty$ a steady state is
reached which decays at the runaway rate $\gamma$ (assuming that a
linear collision operator is employed).  Because $f$ and all its moments
decay at the same rate, $\gamma$ is given from \eqref{cons-grid} as
	$$\gamma={\flux 1\over \numint 1},\Eqlab{gamma-def}$$
which we will take to be the definition of $\gamma$ for all $t$.
Thus we write 
	$$f({\rm v},t)=f'({\rm v},t)
\exp\biggl(-\int_0^t\gamma(t')\, dt'\biggr),\Eqlab{f-decay}$$
	where $\gamma$ is given by \eqref{gamma-def} and
$f'(t\rightarrow\infty)$ is independent of $t$.
	If \eqref{f-decay} is substituted into \eqref{matrix-eq}, we
obtain
	$${\partial f'\over \partial t}+ (A-\gamma)f'=0,\Eqlab{matrix-mod}$$
	where for simplicity we set the inhomogeneous term $h$ to zero.
Because $\gamma$ is expressed as an integral over $f$
\eqref{gamma-def}, it varies slowly and need not be evaluated very
often.  Thus \eqref{matrix-mod} may be regarded as a linear equation and
solved in precisely the same way as \eqref{matrix-eq} (with $h=0$)
except that $\gamma$ must be subtracted from $b_{v,i+1/2,j+1/2}$
eq.~(\ref{mat-coeff}b).

As an example, \figref{run-fig} shows the steady-state distribution
obtained by this method with $Z_i=1$,
$E=0.06$, $M=N=100$, $v_{\rm max}=10$, and
electron-electron collisions given by $C_{\rm Max}^{e/e}$.  Since there
is no rf diffusion term, there are no cross-derivative terms and the
steady state is most easily obtained by taking a constant time step of
$\Delta t=1$.  The runaway rate $\gamma$ is recomputed every ten time
steps and the convergence condition $R=10^{-9}$ is met after 820 time
steps.  In the steady state, we have $\gamma=5.211\times10^{-5}$ and
$J=0.3133$.  These are close to the results obtained by Kulsrud {\sl et
al.}\ \cite{Kulsrud}, namely $\gamma=5.411\times10^{-5}$ and
$J=0.3143$.

Again, it is important to explore the possible errors in these figures.
Extending the boundary to $v_{\rm max}=20$ and doubling $N$ to $200$
gives $\gamma=5.210\times10^{-5}$ and $J=0.4514$.  While there is
practically no change in $\gamma$, $J$ is about $50\%$ larger.  This
discrepancy arises because there is a large contribution to the total
current by the runaways in the region $10<v<20$.  We can verify this by
estimating the total current for an arbitrary $v_{\rm max}$ on the
basis of the results from $v_{\rm max}=10$.  For simplicity, assume that
all the runaways are concentrated near $v_\perp=0$.  From small $\gamma$
and in the limit $t\rightarrow\infty$, the runaway distribution is
independent of $v_\parallel$, so that $f(v_\parallel\gg v_t)\approx
(\gamma/E) \delta({\bf v}_\perp)$.  The current obtained by integrating
$v_\parallel f$ out to $v=v_{\rm max}$ is then
	$$J(v_{\rm max})\approx J_{\rm bulk}
+\half(\gamma/E)v_{\rm max}^2,$$
	where, using the data from $v_{\rm max}=10$, we have $J_{\rm
bulk}=0.270$.  We can interpret $J_{\rm bulk}$ as the current carried by
the bulk electrons and the other term as the current carried by the
runaways.  This now gives $J(v_{\rm max}=20)=0.444$ which is within
$2\%$ of the observed value.  The lesson from this exercise is that it
makes little sense to quote the result for $J$ when the runaway rate is
appreciable because it depends strongly on $v_{\rm max}$.  It is
preferable to determine the bulk current since this is then weakly
dependent on $v_{\rm max}$ and has a physical interpretation.  We have
seen that $v_{\rm max}=10$ is sufficiently large to give $\gamma$ and
$J_{\rm bulk}$ accurately.

In order to determine the effect of the finite mesh on the runaway
results, we vary $M$ and $N$ with $M=N$ and $v_{\rm max}=10$.  The
results for $\gamma$ and $J$ are shown in \figref{conv-e}.  The
asymptotic values are $\gamma=5.185\times10^{-5}$ and $J=0.31334$.  The
errors in the values for $M=N=100$ are $0.5\%$ and $0.02\%$,
respectively.  The errors are considerably less than with the rf
problem in \figref{conv-d} and the convergence is much more
regular (as $\Delta v^2$).

A disadvantage of solving for the decaying steady state of the
distribution, \eqref{matrix-mod}, is that ${\bf S}$ is no longer
divergence free.  This means that the stream lines cannot be plotted as
contours of a stream function $A$, \eqref{stream-grid}.  This can be
remedied by injecting electrons at the origin to match the runaway loss
of particles.  Although this is a rather artificial problem, there is
little error in the runaway rate provided that the runaway rate itself
is small.  We implement this procedure as follows:  The loss of
particles at $v=v_{\rm max}$ is
	$$n\gamma = \flux 1.$$
	We match this loss by a uniform radial flux at the origin
	$$v_0^2 S_{v,i+1/2,0} = {n\gamma \sin(\half\Delta\theta)
\over 2\pi\,\Delta\theta},$$
	which is chosen to give
	$$\sum_{i=0}^{M-1}
2\pi\sin\theta_{i+1/2} v_0^2 S_{v,i+1/2,0}\,\Delta\theta = n\gamma.$$
	(The product $v_0^2 S_{v,i+1/2,0}$ is finite even though
$S_{v,i+1/2,0}$ is infinite.)  From \eqref{flux-grid}, we see that this
introduces a source term
	$v_0^2 S_{v,i+1/2,0}/(v_{1/2}^2\,\Delta v)$ 
	into the expressions for $\partial f_{i+1/2,1/2}/\partial t$.
This is included as part of the inhomogeneous term $h$ in
\eqref{matrix-eq}.  The expressions for the stream function
\eqsref{stream-grid} require a slight modification to give
	$$\eqalign{A_{i,j}&=-\gamma +{2\pi v_j^2\over n}\sum_{i'=0}^{i-1}
\sin\theta_{i'+1/2}S_{v,i'+1/2,j}\,\Delta\theta,
\cr
&=A_{i,0}-{2\pi\sin\theta_i\over n}\sum_{j'=0}^{j-1}
v_{j'+1/2}S_{\theta,i,j'+1/2}\,\Delta v,
\cr}$$
	where the integration constant has been chosen to given
$A_{0,j}=-\gamma$ and $A_{M,j}=0$.

The flux plot computed by this method for the case shown in
\figref{run-fig}, i.e., for $Z_i=1$, $E=0.06$, $M=N=100$, $v_{\rm
max}=10$, is shown in \figref{run-flux}.  When computed in this way, the
runaway rate is slightly lower $\gamma=5.148\times10^{-5}$ because a
typical runaway particle has to be accelerated from $v=0$ instead of
$v=1$.  The current $J=0.3127$ is also lower.

\subsection{Other methods}

An {\sl infinite} time step can be used if the Crank--Nicholson scheme,
eq.\ (\ref{crank}), is modified so that $f^{k+1}$ is used in place of
$\half(f^{k+1}+f^k)$.  Then, the steady state can be achieved in a
single time step.  Of course, this entails inverting the large matrix $A$
(which is why we advocated using the alternating-direction-implicit
method in preference to the Crank--Nicholson method).  However, routines
are available to perform such an inversion and they have been employed
by O'Brien {\sl et al.} \cite{O'Brien}.  An important feature of this
method is the use of disk files to hold intermediate results.
(Typically, the full matrix cannot fit into memory.)  They report a {\sc
CPU} time of 35 s to invert the matrix arising from the discretization
of the Fokker--Planck equation on a $300\times100$ grid with this time
scaling as $MN\times\min(M,N)$.  This method is therefore comparable (as
far as {\sc CPU} time goes) to the Chebyshev acceleration method.  There
are two potential drawbacks of this scheme: Firstly, there is a
significant cost in {\sc I/O} time with this method because of the use
of disk files for storage.  Secondly, the advantage of the method is
reduced if the steady state cannot be reached in a single time step.
This is the case with the more complicated collision operators, because
the matrix $A$ is a function of time.

Various iterative methods are available for obtaining a steady-state
solution \cite{Marchuk}.  These are basically approximate methods of
inverting the matrix $A$.  Notable is Gauss--Seidel relaxation in which
the elements of $f$ are successively updated to achieve $\partial
f/\partial t=0$ at the point in question.  In line relaxation, a whole
line of elements (for example, $j=\rm const$) is updated simultaneously
(requiring the solution of a tridiagonal system of equations).  Line
relaxation gives the same convergence rate as Gauss--Seidel relaxation
and may be vectorized if the even-numbered rows ($j=\rm even$) are
updated in one sweep followed by the update on the odd-numbered rows.

The odd-even line relaxation method is extended with the
successive-over-relaxation method where the over-relaxation parameter
$\omega$ determines how much overshoot there is beyond the value of $f$
which gives $\partial f/\partial t=0$.  Unfortunately, these methods
give results which are roughly the same as using fixed time
steps.  For the example shown in \figref{f-steady}, with the
over-relaxation parameter set to $\omega=1.4$, the convergence
criterion is met after 5980 steps. (Compare this to the 4060 steps
required in the fixed-time-step method.  However, one relaxation step
tends to be computationally less expensive than one step of the
alternating-direction-implicit method.)  For this example, the method
becomes unstable with $\omega\ge1.5$.

Although by themselves relaxation methods are not very useful for this
problem, they are an important ingredient in the multigrid method
\cite{Brandt,multigrid}.  In this method, the problem is solved at
several different grid spacings (usually differing from each other by a
factor of two).  A few relaxation sweeps are carried out on the finest
grid.  Because relaxation is a local method, this is very effective at
damping the short wavelength modes (with wavelength comparable to grid
spacing).  If relaxation is continued on the finest grid, convergence
would become slower because longer wavelength modes would dominate the
residue.  However, in the multigrid method, the residue is transferred
onto the next coarsest grid where relaxation methods are again
efficient.  This process continues recursively up to very coarse grids
where either relaxation methods or direct solution methods can be used.

This method has not been implemented for the Fokker--Planck equation.
However, we can estimate the time required to obtain a steady state.
Each relaxation step on the finest grid gives a reduction in $R$ by
about a factor of two.  (The total work at the coarser grids is at most
a multiple of the work on the finest grid.)  In contrast, the mean
reduction in $R$ with the Chebyshev method is by $4\%$ per step (see
\figref{residu-fig}).  Thus the multigrid method will require about
$\log(0.5)/\log(0.96)\approx 16$ times fewer steps---an
order-of-magnitude improvement over the Chebyshev method.

\section{Relativistic Treatment}\label{relativity}

\subsection{The Fokker--Planck equation}

Fokker--Planck methods have been used to study current drive by
lower hybrid waves.  In a fusion plasma, these waves will interact with
electrons that travel at close to the speed of light.  In such cases, it
is necessary to reformulate the equation to include relativistic
effects.  The first change is that the electron distribution function
is expressed in momentum rather than velocity space so that
\eqref{fp-1} becomes
	$${\partial f_e\over\partial t}-\sum_s C(f_e,f_s) +
\nabla\cdot{\bf S}_w+q_e{\bf E}\cdot\nabla f_e = 0, \Eqlab{fp-rel}$$
	where now the $\nabla\equiv\partial/\partial{\bf p}$ operator
operates in {\sl momentum} space, ${\bf S}_w$ is the rf-induced flux in
momentum space, and $f_e$ is normalized so that
	$$\int f_e({\bf p})\,\dvp=n_e.$$
	In spherical coordinates we have
	$$\nabla\cdot{\bf S}
={1\over p^2}{\partial\over\partial p}p^2S_p+
{1\over p\sin\theta}{\partial\over\partial \theta}\sin\theta S_\theta,$$
	where $\cos\theta=p_\parallel/p$.

In addition, the forms of the collision term and the quasilinear
diffusion term are altered.

\subsection{The relativistic collision operator}

The relativistic collision operator is given by Beliaev and Budker
\cite{Beliaev}.  It can again be written as the divergence of a flux
$C(f_a,f_b)=-\nabla\cdot{\bf S}_c^{a/b}$, where now we have 
	$${\bf S}_c^{a/b}={q_a^2 q_b^2 \over 8\pi\epsilon_0^2}
\ln \Lambda^{a/b}\int \mat U({\bf u})\cdot
\biggl(
f_a({\bf p}) {\partial f_b({\bf p}')\over\partial {\bf p}'}-
f_b({\bf p}') {\partial f_a({\bf p})\over\partial {\bf p}}
\biggr) \,\dvp'.\Eqlab{coll-rel}$$
	The expression for $\mat U$ is rather complicated \cite{Beliaev}.
However, if either the test or the background species is weakly
relativistic ($p\ll m_a c$ or $p'\ll m_b c$), then $\mat U$ may be
approximated by its nonrelativistic form
	$$
\mat U({\bf u})={u^2\mat I - {\bf uu}\over u^3},\qquad
{\bf u}={\bf v}_a-{\bf v}'_b,$$
	where ${\bf v}_s={\bf p}/m_s\gamma_s$ is the velocity of
species $s$, $\gamma_s=(1+p^2/m_s^2c^2)^{1/2}$ is the relativistic
correction factor, and $m_s$ is the rest mass.

Despite the resemblance of \eqref{coll-rel} to \eqref{coll-land}, this
collision operator cannot be readily expressed in terms of Rosenbluth
potentials.  However, considerable progress can still be made by working
directly with \eqref{coll-rel}.  We restrict our attention to
electron-ion and electron-electron collisions.

For collisions off infinitely massive ions, we can take the
ions to be stationary $v_i'\rightarrow0$ and evaluate the integrals to
give
	$$C^{e/i}\bigl(f_e({\bf p})\bigr)=
\Gamma^{e/e}{Z_i\over 2v_ep^2}{1\over\sin\theta}{\partial\over\partial\theta}
\sin\theta{\partial\over\partial\theta}f_e({\bf v}),\Eqlab{coll-rel-ions}$$
where
$$\Gamma^{a/b}={n_b q_a^2 q_b^2 \ln \Lambda^{a/b}\over 4\pi
\epsilon_0^2}$$
(this differs by a factor of $m_a^2$ from the definition given in
\secref{prelim}).

For electron-electron collisions we start with the case of an isotropic
background $C\bigl(f_e({\bf p}),\penalty0 f_e^{(0)}(p)\bigr)$.  The fluxes for
this term are \cite{relpap}
	$$\eqalignno{S_{cp}^{e/e}&=
-D_{cp p}^{e/e} {\partial f_e\over\partial p}
+F_{cp}^{e/e}f_e,&(\eqlab{isotropic-rel}a)\cr
S_{c\theta}^{e/e}&=
-D_{c\theta \theta}^{e/e}{1\over p}{\partial f_e\over\partial \theta}
,&(\ref{isotropic-rel}b)\cr}$$
	where
	$$\eqalignno{
D_{cpp}^{e/e}&={4\pi \Gamma^{e/e}\over 3n_e}
\biggl(\int_0^p p'{}^2f_e^{(0)}(p') {v_e'{}^2\over v_e^3}\,dp'+
\int_p^\infty p'{}^2f_e^{(0)}(p') {1\over v_e'}\,dp'\biggr),
&(\eqlab{iso-rel-gen}a)\cr
D_{c\theta\theta}^{e/e}&={4\pi \Gamma^{e/e}\over 3n_e}
\biggl(\int_0^p p'{}^2f_e^{(0)}(p') {3v_e^2-v_e'{}^2\over 2v_e^3}
\,dp'+
\int_p^\infty p'{}^2f_e^{(0)}(p') {1\over v_e'}\,dp'\biggr),
&(\ref{iso-rel-gen}b)\cr
F_{cp}^{e/e}&=-{4\pi \Gamma^{e/e}\over 3n_e}
\biggl(\int_0^p p'f_e^{(0)}(p') {3v_e'-v_e'{}^3/c^2\over v_e^2}
\,dp'+
\int_p^\infty p'f_e^{(0)}(p') 2v_e/c^2\,dp'\biggr).\qquad
&(\ref{iso-rel-gen}c)\cr
}$$
	These should be compared with their nonrelativistic
counterparts \eqsref{isotropic} and (\ref{iso-gen}).

In the relativistic limit, the Maxwellian distribution \eqref{max}
becomes \cite{DeGroot} 
	$$f_{em}(p)={n_e\over4\pi m_e^2cT_eK_2(\Theta^{-1})}
\exp\biggl(-{{\cal E}\over T_e}\biggr),\eqn(\eqlab{rel-maxwell})$$
	where
	$${\cal E}=m_ec^2\gamma_e$$
is the total electron energy,
	$$\Theta=T_e/m_ec^2=T_e/511\,\rm keV,$$
and $K_n$ is the $n$th-order modified Bessel function of the
second kind.  If we substitute $f_e^{(0)}(p)= f_{em}(p)$ into
\eqsref{iso-rel-gen}, we obtain $F_{cp}^{e/e}/D_{cpp}^{e/e} =-v_e/T_e$.
Thus we find that $f_{em}$ annihilates the electron-electron collision
term $C(f_{em},\linebreak[0]f_{em})=0$.  The integrals in \eqref{iso-rel-gen} cannot
be performed analytically with $f_e^{(0)}(p)= f_{em}(p)$ and so in the
numerical code these are performed numerically.

For the Maxwellian distribution \eqref{rel-maxwell}, we define a
thermal momentum
	$$p_{te}=\sqrt{m_eT_e},$$
a thermal velocity
$$v_{te}^2={1\over 3n_e}\int v_e^2
f_{em}(p)\,\dvp={T_e\over m_e}\biggl(
1-{5\over2}\Theta+
{55\over8}\Theta^2+\ldots\biggr),$$
and a thermal collision frequency
$$\nu_{te}={m_e\Gamma^{e/e}\over p_{te}^3}.$$

For $p\gg p_{te}$, the indefinite limits in the integrals
in Eq.~(\ref{iso-rel-gen}) can be replaced by $\infty$, giving
\cite{Mosher}
$$\eqalignno{D_{cpp}^{e/e}
&=\Gamma^{e/e} {v_{te}^2\over v_e^3},&(\eqlab{rel-high}a)\cr
D_{c\theta\theta}^{e/e}
&=\Gamma^{e/e}{1\over 2v_e}\biggl(1-{v_{te}^2\over v_e^2}\biggr),
&(\ref{rel-high}b)\cr
F_{cp}^{e/e}&=-\Gamma^{e/e} {v_{te}^2\over T_ev_e^2}.&(\ref{rel-high}c)\cr
}$$
	These should be compared with \eqsref{iso-high}.

For a background which consists of just the first Legendre harmonic,
the collision term is $C\bigl(f_{em}(p),\penalty100 f_e^{(1)}(p)
\cos\theta\bigr)$.  This is given by \cite{relpap}
	$$\eqalignno{{C\bigl(f_{em}(p), f_e^{(1)}(p)\cos\theta\bigr)
\over f_{em}(p)\cos\theta}\kern-7em&\kern7em
={4\pi \Gamma^{e/e}\over n_e}\times \cr
\Biggl\{&{m_e f_e^{(1)}(p) \over \gamma_e}
+{1\over 5}\int_0^p p'{}^2f_e^{(1)}(p'){m_e\over T_e}\biggl[
{\gamma_e\over p^2}{v_e'\over\gamma_e'{}^3}
\biggl({T_e\over m_ec^2}(4\gamma_e'{}^2+6)
-{1\over3}(4\gamma_e'{}^3-9\gamma_e')\!\biggl)\cr
&\qquad\qquad\qquad\qquad\qquad\qquad\qquad
{}+{\gamma_e^2\over p^2}{v_e'\over\gamma_e'{}^3}
\biggl({m_ev_e'{}^2\over T_e}\gamma_e'{}^3
-{1\over3}(4\gamma_e'{}^2+6)\!\biggl)
\biggr]\,dp'\cr
&\qquad\qquad{}+{1\over 5}\int_p^\infty p'{}^2f_e^{(1)}(p'){m_e\over T_e}\biggl[
{\gamma_e'\over p'{}^2}{v_e\over\gamma_e^3}
\biggl({T_e\over m_ec^2}(4\gamma_e^2+6)
-{1\over3}(4\gamma_e^3-9\gamma_e)\!\biggl)\cr
&\qquad\qquad\qquad\qquad\qquad\qquad\qquad
{}+{\gamma_e'{}^2\over p'{}^2}{v_e\over\gamma_e^3}
\biggl({m_ev_e^2\over T_e}\gamma_e^3
-{1\over3}(4\gamma_e^2+6)\!\biggl)
\biggr]\,dp'\Biggr\}.&(\eqlab{rel-first-lin})\cr}$$
	[Compare with \eqref{first-lin}.]  The general solution of the
linearized electron-electron collision operator $C(f_e,f_{em}) +
C(f_{em},f_e)=0$ is
	$$f_e=(a+{\bf b}\cdot{\bf p}+d{\cal E})f_{em},$$
where $a$, $\bf b$, and $d$ are arbitrary constants.  With $a=d=0$ and
${\bf b}=\hat{\bf p}_\parallel$, this provides a
useful check on Eqs.~(\ref{iso-rel-gen}) and (\ref{rel-first-lin}) and their
computational realizations.

In the example we show below, we use the electron-ion collision
operator given by \eqref{coll-rel-ions} and the relativistic
generalization of the truncated collision operator \eqref{truncated}
	$$C_{\rm trunc}^{e/e}\bigl(f_e({\bf p})\bigr)=
C\bigl(f_e({\bf p}),f_{em}(p)\bigr)
+C\bigl(f_{em}(p), f_e^{(1)}(p)\cos\theta\bigr),
\Eqlab{rel-truncated}$$
	where the first term is given by \eqsref{isotropic-rel} and
(\ref{iso-rel-gen}) and the second term by \eqref{rel-first-lin}.

\subsection{Wave-particle interaction}

We saw in \secref{ql} that the quasilinear diffusion operator had two
principal ingredients:  the wave-particle resonance condition, and the
diffusion paths.  Both of these are modified by relativistic effects.

The wave-particle resonance condition becomes
	$$\omega-k_\parallel v_{e\parallel}-n\Omega_e/\gamma_e=0,$$
where $\Omega_e=q_eB/m_e$ is the rest-mass cyclotron frequency.
Translating this into momentum space gives
	$$\omega\sqrt{1+p^2/m_e^2c^2}-k_\parallel p_\parallel-n\Omega_e=0.$$
	This modification of the resonance condition is important in
the consideration of current drive by electron cyclotron waves
\cite{Cairns}.

The diffusion paths are again given by surfaces of constant energy in
the wave frame.  The expression for the energy in a frame moving at
$(\omega/k_\parallel){\bf \hat p}_\parallel$ is 
	$${\cal E}'={{\cal E}-(\omega/k_\parallel)p_\parallel
\over\sqrt{1-\omega^2/k_\parallel^2c^2}}.$$
The diffusion paths are, therefore, given by
	$${\cal E}-(\omega/k_\parallel)p_\parallel=\rm const.$$
	These paths are parallel to the vector
	$$\biggl({\omega\over k_\parallel}-v_{e\parallel}\biggr)
{\bf \hat p}_\perp+v_{e\perp}{\bf \hat p}_\parallel.$$
	This should be compared with the vector ${\bf a}_n$ defined in
\secref{ql}.  The paths are ellipses or hyperbolae in momentum space
depending on whether $\omega/k_\parallel$ is less than or greater than
$c$ \cite{Karney-ec}.

For lower hybrid waves, we have $n=0$ and the diffusion is
in the parallel direction.  We, therefore, generalize \eqsref{d-ql}
by incorporating the modified resonance condition to read
	$$\mat D_w=D_w(p_\perp,p_\parallel)
{\bf\hat p}_\parallel{\bf\hat p}_\parallel,\eqn(\eqlab{d-ql-rel}a)$$
	where
	$$D_w(p_\perp,p_\parallel)=\cases{
D_0,&for $v_1<p_\parallel/\gamma_e<v_2$,\cr
0,&otherwise.\cr}\eqn(\ref{d-ql-rel}b)$$

\subsection{Example}

To illustrate the relativistic effects we show in \figref{rel-fig} the
steady-state distribution function obtained by integrating the
Fokker--Planck equation with electron-electron collisions given by
$C_{\rm trunc}^{e/e}$ \eqref{rel-truncated} and electron-ion collisions
given by $C^{e/i}$ \eqref{coll-rel-ions} with $Z_i=1$.  The quasilinear
diffusion term is given by \eqref{d-ql-rel} with $D_0=1$, $v_1=0.4c$,
and $v_2=0.7c$.  (Except for the perpendicular profile of $\mat D_w$,
this is the same as the example given in \refref{relpap}.)  The
integration is carried out with $M=N=100$ and $p_{\rm max}=20$.  We
normalize all momenta to $p_{te}$, velocities to $p_{te}/m_e$ ({\sl
not} $v_{te}$), the current density to $n_eq_ep_{te}/m_e$, the power
density to $n_ep_{te}^2\nu_{te}/m_e$, etc.  Again we are principally
interested in the current and the power dissipated.  These are defined
by
	$$\eqalign{J&=
{\numint (v_e\cos\theta)\over n},\cr
P&={1\over n}\sum_{i=0}^{M-1}\sum_{j=0}^{N}
2\pi\sin\theta_{i+1/2} v_jp_j^2 S_{wp,i+1/2,j} \,\Delta p\Delta\theta,\cr}$$
	where $\numint$ is the generalization of \eqref{num-int} to
momentum space, $n=\numint 1$.  [Compare these expressions with
\eqsref{curr-def-grid} and (\ref{pd-def-grid}).]  In the steady state,
we find $J=3.732\times10^{-3}$, $P=1.256\times10^{-4}$, and
$J/P=29.72$.

Again a useful benchmark is provided by the electrical conductivity.  In
the limit $E\rightarrow0$ this is correctly given if $C_{\rm
trunc}^{e/e}$ is employed.  With $E=10^{-3}$, $Z_i=1$, $\Theta=0.01$,
$M=N=100$, and $p_{\rm max}=10$, we find $J/E=7.307$, which differs from
the true value of 7.291 by about $0.2\%$.  Values of the conductivity
for various values of $Z_i$ and $\Theta$ are tabulated in
\tabref{rel-conduct}.

\section{Adjoint Method}\label{adjoint}

\subsection{Introduction and example}

We have considered here techniques for solving the Fokker--Planck
equation with an added quasilinear diffusion term.  This tends to be an
expensive operation because the addition of the quasilinear diffusion
term greatly increases the parameter space to be scanned.  For example,
the study of lower hybrid current drive \cite{Karney-lh} included the
results of some 50 runs with different values of $v_1$ and $v_2$.  Even
so, no systematic study was made of the dependence on the
parameters $D_0$ and $Z_i$.

However, the amount of work can be drastically reduced using the
adjoint method.  This was introduced by Hirshman \cite{Hirshman} for the
study of beam-driven currents.  Later, Antonsen and Chu \cite{Antonsen}
used it to study rf-driven currents.

To illustrate the method, we will outline the analysis given by
Antonsen and Chu \cite{Antonsen}.  The method begins by assuming that
$f_e$ is close to a Maxwellian $f_{em}$ so that the linearized
electron-electron collision operator $C_{\rm lin}^{e/e}$ \eqref{linear}
can be used.  The quasilinear diffusion term is taken as a {\sl given}.  As
pointed out in \secref{ql}, the Fokker--Planck equation then becomes an
inhomogeneous equation, whose linear operator is independent of the
wave drive.  Two further assumptions are made, namely that ${\bf E}=0$
and that a steady state has been reached.  (Neither of these assumptions
is necessary and they have been relaxed in \refref{Fischa}.)  The
Fokker--Planck equation is then
	$$\eqalignno{C\bigl(f_e({\bf v})\bigr)&
\equiv C_{\rm lin}^{e/e}\bigl(f_e({\bf v})\bigr)
+C^{e/i}\bigl(f_e({\bf v})\bigr)\cr
&=\nabla\cdot {\bf S}_w+ \biggl({m_ev^2\over
2T_e}-{3\over 2}\biggr)f_{em}(v) {\partial \ln T_e\over \partial t},
&(\eqlab{Chap-eq})\cr}$$
	where we have inserted the Chapman--Enskog--Braginskii energy
loss term to ensure that \eqref{Chap-eq} has a solution (i.e., to
ensure that the Fokker--Planck equation reaches a steady state).  Taking
the energy moment of this equation, and noting that the collision
operator is energy conserving, we find the equation for $\partial
T_e/\partial t$
	$${\partial\over \partial t}\biggl({3\over2}n_eT_e\biggr)=P,$$
	where $P$ is given by \eqref{pd-def}.

The straightforward approach is now to solve \eqref{Chap-eq} for a
particular ${\bf S}_w$, determine the electron distribution $f_e$, and hence
find the rf-driven current.  The adjoint method gives a way of computing
the current without having to find $f_e$.  Consider first the ``adjoint''
problem 
	$$C\bigl(f_{em}(v)\chi({\bf v})\bigr)=-q_ev_\parallel f_{em}(v),
\Eqlab{adjoint-eq}$$
	where we require that $f_{em}\chi$ have zero density and zero energy.
This is the Spitzer--H\"arm equation for the perturbed distribution in
the presence of an electric field ${\bf E}=T_e{\bf \hat v}_\parallel$.
Let us multiply \eqref{adjoint-eq} by $f_e/f_{em}$ and integrate over
velocity. This gives
	$$J=-\int (f_e/f_{em}) C(f_{em}\chi)\,\dv,$$
where $J$ is the current carried by the electron distribution $f_e$.
Now we utilize the self-adjointness of the linearized collision
operator
	$$\int\psi C(f_{em}\chi)\,\dv=\int\chi C(f_{em}\psi)\,\dv,$$
	together with \eqref{Chap-eq} for $C(f_e)$ to give
	$$J=\int {\bf S}_w({\bf v})\cdot\nabla\chi({\bf v})\,\dv.
\Eqlab{adjoint-curr}$$
	\Eqref{adjoint-curr} is the desired expression for the current.
The quantity $\chi$ serves as the Green's function for the current $J$.
The current drive efficiency is given by
	$${J\over P} ={\int {\bf S}_w\cdot\nabla\chi\,\dv\over \int
m_e{\bf S}_w\cdot{\bf v}\,\dv}.\Eqlab{adjoint-eff}$$

\subsection{Solving the adjoint equation}

In order to apply this method, we must determine $\chi$ by solving
\eqref{adjoint-eq}.  Because $\chi({\bf v})$ consists of only the first
Legendre harmonic $\chi^{(1)}(v)\cos\theta $, this equation reduces to a
one-dimensional integro-differential equation,
	$${1\over v^2}{\partial\over\partial v}
v^2 D_{cvv}^{e/e}{\partial\chi^{(1)}\over\partial v}
-{m_e v\over T_e}D_{cvv}^{e/e}{\partial\chi^{(1)}\over\partial v}
-{2D_{c\theta\theta}^{e/e}+\Gamma^{e/e}Z_i/v\over v^2}\chi^{(1)}
+I^{e/e}(\chi^{(1)})+q_ev=0,\eqn(\eqlab{adjoint1})$$
	where $D_{cvv}^{e/e}$ and $D_{c\theta\theta}^{e/e}$ are given by
\eqsref{iso-max}, and $I^{e/e}$ is defined by
	$$I^{e/e}(\chi^{(1)}(v))=
{C(f_{em}(v),f_{em}(v)\chi^{(1)}(v)\cos\theta )\over f_{em}(v)\cos\theta }$$
	[see \eqref{first-lin}].

In general, \eqref{adjoint1} must be solved numerically.  This is done
by constructing the partial differential equation by setting the
left-hand side of \eqref{adjoint1} equal to ${\partial
\chi^{(1)}/\partial t}$.  The resulting equation is integrated in time
with arbitrary initial conditions until a steady state is reached.  The
integration is carried out in the domain $0<v<v_{\rm max}$ and the
boundary conditions are taken to be $\chi^{(1)}(v=0)=0$ and
$\partial^2\chi^{(1)}(v=v_{\rm max})/\partial v^2=0$.

Approach to the steady state is accelerated by treating the first three
terms in \eqref{adjoint1} fully implicitly; i.e., in order to compute
$[\chi^{(1)}(t+\Delta t)-\chi^{(1)}(t)]/\Delta t$, we evaluate these terms at
$t+\Delta t$.  This means that very large time steps can be used.  The
integral term $I^{e/e}(\chi^{(1)})$ is treated explicitly and is
reevaluated at each time step.  The resulting difference equations form
a tridiagonal matrix which can be solved by Gaussian elimination.

Because the adjoint equation is the same as the equation for the perturbed
distribution in the presence of a weak electric field, we can solve
\eqref{adjoint1} to obtain values of the electrical conductivity which
is defined by
	$${J\over E}=\int {q_e v_\parallel\over T_e}f_{em}(v)\chi({\bf v})\,\dv
={4\pi q_e\over3T_e}\int v^3 f_{em}(v)\chi^{(1)}(v)\,dv.$$
	This procedure was carried out using the method outlined above
with $v_{\rm max}=15 v_{te}$, $\Delta v=0.001 v_{te}$, and $\Delta
t=1000/\nu_{te}$.  Because we are only working with a one-dimensional
equation, it is possible to use a much finer mesh than with
two-dimensional problems and so obtain results which are effectively
``exact.''  The results are summarized in \tabref{conduct} where we have
also included the results from use of approximate collision operators.
The same technique is easily generalized to relativistic plasmas using
the collision operator given in \secref{relativity}.  This gives the
relativistic corrections to the conductivity which are given in
\tabref{rel-conduct}.

When the adjoint method is applied to more complicated situations
(e.g., including a dc electric field), a two-dimensional equation must
be solved.  We can then use many of the techniques for the solution of
the Fokker--Planck equation, which have been presented in the preceding
sections.

\subsection{Discussion}

Let us assess the work involved in utilizing the adjoint method.  Once
the adjoint equation has been solved, the current and the efficiency
are immediately given in terms of ${\bf S}_w$ by \eqsref{adjoint-curr}
and (\ref{adjoint-eff}).  Instead of having to solve the Fokker--Planck
equation afresh for every form of ${\bf S}_w$, a couple of velocity
integrals over ${\bf S}_w$ suffice to give the important quantities.
The parameter space that must be scanned in order to give a complete
understanding of the physics is greatly reduced.  The adjoint method
does not give the electron distribution $f_e$ nor the rf-induced flux
${\bf S}_w$.  On the other hand, a crude estimate of ${\bf S}_w$ gives
an accurate estimate of the efficiency because \eqref{adjoint-eff}
involves the ratio of two integrals over ${\bf S}_w$.  An effective way
to use this method within a ray-tracing code would be to determine
${\bf S}_w$ from a solution of the one-dimensional Fokker--Planck
equation \cite{Fisch1} and to use this to determine both $J$ and $P$
from \eqsref{adjoint-curr} and (\ref{pd-def}).  The code thereby
benefits from an accurate determination of the current drive efficiency
while the high computational costs of integrating the two-dimensional
Fokker--Planck equation are avoided.

Because the current drive efficiency is determined by a single function
$\chi$, it is possible to ask questions not readily answerable from
numerical solutions of the Fokker--Planck equation.  Examples are: What
is the asymptotic form for the efficiency as the wave phase velocity
becomes large?  What is the maximum possible efficiency for a particular
class of waves?

Besides determining the current, the adjoint method can be adapted to
give other moments of the electron distribution by changing the
right-hand side of \eqref{adjoint}.  This can then give, for example,
the perpendicular energy of the electrons, bremsstrahlung radiation,
etc.  This method has been used to determine the current-drive
efficiency in a relativistic plasma \cite{relpap}.  Recent developments
of the method \cite{Fischa} allow the determination of arbitrary
moments of $f_e$ (not just the current $J$), and the determination of
the time development of such moments.  These have been applied to the
study of rf current ramp-up \cite{ramp-pap}.

\section{Conclusions}

In the last fifteen years, Fokker--Planck codes have gone from esoteric
programs developed by a few researchers which could only be run on a
few machines to widely available tools used by a large number of
physicists on many different computers.  This has been due to the large
increase in computer power available to the average physicist and the
pioneering efforts of Killeen {\sl et al.}\ \cite{Killeen1,Killeen2}.

In this paper, I have given a detailed description of a particular
implementation of a code to solve the Fokker--Planck equation with
emphasis on a particular application, namely current drive by
lower hybrid waves.  There are many other implementations of this code
that have been applied to a large variety of interesting problems.
My goal has been to illustrate the main numerical problems by means of
concrete examples.  The methods presented here cover the major numerical
problems that are encountered in all Fokker--Planck codes.

There are two areas which still require attention.  Firstly, improved
methods for obtaining the steady-state solution of the Fokker--Planck
equation are needed.  Here the multigrid method offers the best promise for
substantial savings over the other methods described in this paper.
Secondly, the adjoint methods outlined in \secref{adjoint} should be
extended and applied to a wider range of problems.  Ray-tracing codes
still need to be modified to accept the results of these calculations.

\section*{Acknowledgments}

I would like to thank N.  J.  Fisch for a very fruitful collaboration
extending over several years on various problems in rf current drive,
which provided the impetus for the work described here.

This work was supported by the United States Department of Energy under
Contract DE--AC02--76--CHO--3073.

\appendix\section{Numerical Techniques}\label{numerical} \input fortran.sty

In this appendix, various fragments of code are shown.  A two-di\-men\-sion\-al
Fokker--Planck code is ideally suited to a vector processing machine
like the Cray--1.  However, care must be taken to order the loops
correctly, otherwise they will not vectorize.

The first example is the computation of the current
$\numint(v\cos\theta)$ \eqref{num-int}.  This illustrates the
rather peculiar way in which {\sc FORTRAN} code must be written in
order to take advantage of the Cray--1's architecture \cite{CFT}.  We
assume that the arrays and variables given in \tabref{fortran-vars}
have been initialized as indicated.
	$$\fortcode{
&	\dimension temp(0:iy-1)		\cr
&	\do 1 i=0,iy-1			\cr
&\1	temp(i)=0.0			\cr
1&	\continue			\cr
&	\do 3 j=0,jx-1			\cr
&\1	\do 2 i=0,iy-1			\cr
&\2	temp(i)=temp(i)+x(j)\^3*f(i,j)	\cr
2&\1	\continue			\cr
3&	\continue			\cr
&	\do 4 i=0,iy-1			\cr
&\1	temp(i)=sn(i)*cn(i)*temp(i)	\cr
4&	\continue			\cr
&	cur=0.0				\cr
&	\do 5 i=0,iy-1			\cr
&\1	cur=cur+temp(i)			\cr
5&	\continue			\cr
&	cur=2.0*pi*dx*dy*cur		\cr
}$$
	The important point is that the inner loop (with label {\it 2})
vectorizes.  This would not happen if the order of the
loops were reversed.  There is no particular advantage in taking the
computation of $\fort x(j)\^3$ out of the inner loop since the {\sc
CFT} compiler does this automatically.  The only loop that the compiler
treats inefficiently is the last one.  In fact, we replace this by a
call to the {\sc OMNILIB} routine ssum.

The second example is computing the integral part of the
truncated collision operator
$C\bigl(f_m(v),\penalty100
f^{(1)}(v)\cos\theta\bigr)/\cos\theta$ \eqref{first-lin}.  Here
again it is easy to arrange so that most of the code vectorizes
\cite{McCoy}.  The computation of this term is then relatively
inexpensive compared with the other computations.
	$$\fortcode{
&	\dimension {\it s0}(-1:jx-1),{\it s3}(0:jx),{\it s5}(0:jx),
				{\it f1}(0:jx-1)		\cr
&	\do 1 j=0,jx-1						\cr
&\1	{\it f1}(j)=0.0						\cr
1&	\continue						\cr
&	\do 3 i=0,iy-1						\cr
&\1	\do 2 j=0,jx-1						\cr
&\2	{\it f1}(j)={\it f1}(j)+1.5*dy*sn(i)*cn(i)*f(i,j)	\cr
2&\1	\continue						\cr
3&	\continue						\cr
&	\do 4 j=0,jx-1						\cr
&\1	{\it s0}(j-1)=dx*{\it f1}(j)				\cr
&\1	{\it s3}(j+1)={\it s0}(j-1)*x(j)\^3			\cr
&\1	{\it s5}(j+1)={\it s3}(j+1)*x(j)\^2			\cr
4&	\continue						\cr
&	{\it s3}(0)=0.5*{\it s3}(1)				\cr
&	{\it s5}(0)=0.5*{\it s5}(1)				\cr
&	\do 5 j=1,jx-1						\cr
&\1	{\it s3}(j)={\it s3}(j-1)+0.5*({\it s3}(j)+{\it s3}(j+1))\cr
&\1	{\it s5}(j)={\it s5}(j-1)+0.5*({\it s5}(j)+{\it s5}(j+1))\cr
5&	\continue						\cr
&	{\it s0}(jx-1)=0.5*{\it s0}(jx-2)			\cr
&	\do 6 j=jx-2,0,-1					\cr
&\1	{\it s0}(j)={\it s0}(j+1)+0.5*({\it s0}(j)+{\it s0}(j-1))\cr
6&	\continue						\cr
&	\do 7 j=0,jx-1						\cr
&\1	{\it c1}(j)=({\it s5}(j)/5.0-{\it s3}(j)/3.0)/x(j)\^2	\cr
\cont&\1\hphantom{{\it c1}(j)=}
	+{\it s0}(j)*(x(j)\^2/5.0-1.0/3.0)*x(j)			\cr
&\1	{\it c1}(j)=4*pi*{\it fm}(j)*({\it f1}(j)+{\it c1}(j))\cr
7&	\continue						\cr
}$$
	All the loops vectorize with the exception of the indefinite
integration loops (with labels {\it 5} and {\it 6}).  Most of the time
is spent in the inner loop {\it 2} during the computation of $f^{(1)}$
\eqref{legend-decomp}.

Finally, we consider vectorized Gaussian elimination.  This subroutine
performs Gaussian elimination for the tridiagonal system of equations
	$$x_{i,j}+ \half\Delta t
(a_{i,j}x_{i-1,j}+b_{i,j}x_{i,j}+c_{i,j}x_{i+1,j})=y_{i,j},$$
	to give $x_{i,j}$ for $0\le i< n$, $0\le j<m$.  The coefficients
satisfy $a_{0,j}=c_{n-1,j}=0$.  A substantial fraction of the running time of
the Fokker--Planck code is spent in this subroutine.  When implemented
for a single system of equations $m=1$, this leads to ``vector
dependencies'' which inhibit vectorization.  The solution is to solve
the $m$ systems in parallel with $j$ being the index for the inner
loops.  In the subroutine below, it is assumed that all the matrices are
the same size, that the spacing in memory between $x_{i,j}$ and
$x_{i+1,j}$ (the solution direction) is $ns$, and that the spacing
between $x_{i,j}$ and $x_{i,j+1}$ (the vectorizing direction) is $ms$.
This subroutine uses $x$ and $y$ as temporary storage; thus the initial
data in $y$ are destroyed.
	$$\fortcode{
&	\subroutine {solve}(x,ns,n,ms,m,a,b,c,y,dt)		\cr
&	\dimension  x(0:ms-1,0:m-1),y(0:ms-1,0:m-1),		\cr
\cont&	\hphantom{\dimension}a(0:ms-1,0:m-1),b(0:ms-1,0:m-1),	\cr
\cont&	\hphantom{\dimension}c(0:ms-1,0:m-1)			\cr
&	{\it dt2}=0.5*dt					\cr
&	\do 2 i=0,n-1						\cr
&\1	ia=ns*(i-1)						\cr
&\1	ib=ns*i							\cr
&\1	\do 1 j=0,m-1						\cr
&\2	den=1.0/(1.0+{\it dt2}*(b(ib,j)+a(ib,j)*y(ia,j)))	\cr
&\2	x(ib,j)=(y(ib,j)-{\it dt2}*a(ib,j)*x(ia,j))*den		\cr
&\2	y(ib,j)=-{\it dt2}*c(ib,j)*den				\cr
1&\1	\continue						\cr
2&	\continue						\cr
&	\do 4 i=n-2,0,-1					\cr
&\1	ib=ns*i							\cr
&\1	ic=ns*(i+1)						\cr
&\1	\do 3 j=0,m-1						\cr
&\2	x(ib,j)=y(ib,j)*x(ic,j)+x(ib,j)				\cr
3&\1	\continue						\cr
4&	\continue						\cr
&	\return							\cr
&	\end							\cr
}$$
	There are a couple of tricky points here.  Firstly, we use
nonstandard indexing into the arrays.  The element $x_{i,j}$ is accessed
by the array element $x(ns*i,j)$.  If $ms=1$, then $ns*i$ will generally
exceed the upper bound $ms-1$ on the first dimension of the arrays.
This type of array indexing may cause problems with compilers that
perform bounds checking.  Secondly, we have utilized the fact that
$a(0)=c(n-1)=0$ and assumed that an arbitrary (possibly undefined)
number multiplied by zero will give zero.  If this is not the case, the
$i=0$ and $i=n-1$ iterations in the loop with label $\it 2$ will have
to be split off from the rest of the loop and treated separately.

This subroutine is sufficiently general to be used for both the
matrix inversions required in implementing \eqref{split}.  Assuming
that all the matrices are dimensioned by, for example,
	$$\fortcode{&\dimension f(0:iyl-1,0:jxl-1)\cr}$$
	then the inversions are obtained by
	$$\fortcode{
&	\call {solve}(xia,iyl,jx,1,iy,ax,bx,cx,phi,dt)\cr
&	\call {solve}(xib,1,iy,iyl,jx,ay,by,cy,xia,dt)\cr}$$

\vspace{0.5in}
\bibliographystyle{aip}\bibliography{fp}

\begin{thebibliography}{10}

\bibitem{Fisch1}
N.~J. Fisch,
\newblock Phys. Rev. Lett. {\bf 41}, 873 (1978).

\bibitem{Killeen1}
J.~Killeen and K.~D. Marx,
\newblock in {\em Methods in Computational Physics}, edited by B.~Alder,
  S.~Fernback, and M.~Rothenberg, volume~9, page 421, Academic, New York, 1970.

\bibitem{Killeen2}
J.~Killeen, A.~A. Mirin, and M.~E. Rensink,
\newblock in {\em Methods in Computational Physics}, edited by B.~Alder,
  S.~Fernback, and M.~Rothenberg, volume~16, page 389, Academic, New York,
  1976.

\bibitem{McCoy}
M.~G. McCoy, A.~A. Mirin, and J.~Kileen,
\newblock Computer Phys. Comm. {\bf 24}, 37 (1981).

\bibitem{Kulsrud}
R.~M. Kulsrud, Y.-C. Sun, N.~K. Winsor, and H.~A. Fallon,
\newblock Phys. Rev. Lett. {\bf 31}, 690 (1973).

\bibitem{Karney-lh}
C.~F.~F. Karney and N.~J. Fisch,
\newblock Phys. Fluids {\bf 22}, 1817 (1979).

\bibitem{Cutler}
T.~A. Cutler, L.~D. Pearlstein, and M.~E. Rensink,
\newblock Computation of the bounce average code,
\newblock Technical Report UCRL--52233, Lawrence Livermore Laboratory, 1977.

\bibitem{Kerbel}
G.~D. Kerbel and M.~G. McCoy,
\newblock Phys. Fluids {\bf 28}, 3629 (1985).

\bibitem{Matsuda}
Y.~Matsuda and J.~J. Stewart, Jr.,
\newblock J. Comput. Phys. {\bf 66}, 197 (1986).

\bibitem{Abramowitz}
M.~Abramowitz and I.~A. Stegun,
\newblock {\em Handbook of Mathematical Functions},
\newblock Dover, New York, 1965.

\bibitem{Landau}
L.~D. Landau,
\newblock Phys. Z. Sowjet. {\bf 10}, 154 (1936).

\bibitem{NRL}
D.~L. Book,
\newblock {\em The NRL Plasma Formulary},
\newblock Naval Research Laboratory, Washington, D.C., 1983.

\bibitem{Rosenbluth}
M.~N. Rosenbluth, W.~M. MacDonald, and D.~L. Judd,
\newblock Phys. Rev. {\bf 107}, 1 (1957).

\bibitem{Trubnikov1}
B.~A. Trubnikov,
\newblock Sov. Phys. JETP. {\bf 7}, 926 (1958).

\bibitem{Trubnikov}
B.~A. Trubnikov,
\newblock in {\em Reviews of Plasma Physics}, edited by M.~A. Leontovich,
  volume~1, page 105, Consultants Bureau, New York, 1965.

\bibitem{Harvey}
R.~W. Harvey, K.~D. Marx, and M.~G. McCoy,
\newblock Nucl. Fusion {\bf 21}, 153 (1981).

\bibitem{Chapman}
S.~Chapman and T.~G. Cowling,
\newblock {\em The Mathematical Theory of Non-uniform Gases},
\newblock Cambridge University Press, Cambridge, 3rd edition, 1970.

\bibitem{Braginskii}
S.~I. Braginskii,
\newblock in {\em Reviews of Plasma Physics}, edited by M.~A. Leontovich,
  volume~1, page 205, Consultants Bureau, New York, 1965.

\bibitem{relpap}
C.~F.~F. Karney and N.~J. Fisch,
\newblock Phys. Fluids {\bf 28}, 116 (1985).

\bibitem{Fisch-Karney}
N.~J. Fisch and C.~F.~F. Karney,
\newblock Phys. Fluids {\bf 24}, 27 (1981).

\bibitem{Spitzer}
L.~Spitzer and R.~H\"arm,
\newblock Phys. Rev. {\bf 89}, 977 (1953).

\bibitem{Kennel}
C.~F. Kennel and F.~Engelmann,
\newblock Phys. Fluids {\bf 9}, 2377 (1966).

\bibitem{Bernstein}
I.~B. Bernstein and D.~C. Baxter,
\newblock Phys. Fluids {\bf 24}, 108 (1981).

\bibitem{Bers}
A.~Bers,
\newblock in {\em Plasma Physics---Les Houches 1972}, edited by C.~DeWitt and
  J.~Peyraud, page 113, {Gordon and Breach}, New York, 1975.

\bibitem{Karney-ec}
C.~F.~F. Karney and N.~J. Fisch,
\newblock Nucl. Fusion {\bf 21}, 1549 (1981).

\bibitem{asymp}
N.~J. Fisch and C.~F.~F. Karney,
\newblock Phys. Fluids {\bf 28}, 3107 (1985).

\bibitem{Kritz}
A.~H. Kritz, K.~Appert, L.~Muschietti, and J.~Vaclavik,
\newblock in {\em Non-Inductive Current Drive in Tokamaks, Proc. IAEA Technical
  Committee Meeting, Culham, England}, edited by D.~F.~H. Start, volume~I, page
  161, 1983.

\bibitem{Succi}
S.~Succi, K.~Appert, W.~Core, H.~Hamn\'en, T.~Hellsten, and J.~Vaclavik,
\newblock Comp. Phys. Comm. {\bf 40}, 137 (1986).

\bibitem{Fuchs}
V.~Fuchs, M.~M. Shoucri, A.~Bers, and R.~A. Cairns,
\newblock Technical Report TV RI 187e, Institut de Recherche d'Hydro-Qu\'ebec,
  1985.

\bibitem{Chang}
J.~S. Chang and G.~Cooper,
\newblock J. Comput. Phys. {\bf 6}, 1 (1970).

\bibitem{Marchuk}
G.~I. Marchuk,
\newblock {\em Methods of Numerical Mathematics},
\newblock Springer--Verlag, New York, 1975.

\bibitem{Hewett}
D.~W. Hewett, V.~B. Krapchev, K.~Hizanidis, and A.~Bers,
\newblock Technical Report PFC/RR--84--18, Massachusetts Institute of
  Technology, Plasma Fusion Center, 1984.

\bibitem{O'Brien}
M.~R. O'Brien, M.~Cox, and D.~F.~H. Start,
\newblock Comp. Phys. Comm. {\bf 40}, 123 (1986).

\bibitem{Brandt}
A.~Brandt,
\newblock Math. Comp. {\bf 31}, 333 (1977).

\bibitem{multigrid}
W.~Hackbusch and U.~Trottenberg, editors,
\newblock {\em Multigrid Methods}, volume 960 of {\em Lecture Notes in
  Mathematics},
\newblock Springer--Verlag, Berlin, 1982.

\bibitem{Beliaev}
S.~T. Beliaev and G.~I. Budker,
\newblock Sov. Phys. Doklady {\bf 1}, 218 (1956).

\bibitem{DeGroot}
S.~R. de~Groot, W.~A. van Leeuwen, and C.~G. van Weert,
\newblock {\em Relativistic Kinetic Theory},
\newblock North--Holland, Amsterdam, 1980.

\bibitem{Mosher}
D.~Mosher,
\newblock Phys. Fluids {\bf 18}, 846 (1975).

\bibitem{Cairns}
R.~A. Cairns, J.~Owen, and C.~N. Lashmore-Davies,
\newblock Phys. Fluids {\bf 26}, 3475 (1983).

\bibitem{Hirshman}
S.~P. Hirshman,
\newblock Phys. Fluids {\bf 23}, 1238 (1980).

\bibitem{Antonsen}
T.~M. Antonsen, Jr. and K.~R. Chu,
\newblock Phys. Fluids {\bf 25}, 1295 (1982).

\bibitem{Fischa}
N.~J. Fisch,
\newblock Phys. Fluids {\bf 29}, 172 (1986).

\bibitem{ramp-pap}
C.~F.~F. Karney and N.~J. Fisch,
\newblock Phys. Fluids {\bf 29}, 180 (1986).

\bibitem{CFT}
Cray Research, Inc.,
\newblock {\em CFT, the Cray--1 FORTRAN Compiler}, 1984.

\end{thebibliography}

\begin{thetables}{99}
\newdimen\digitwidth\setbox0=\hbox{\rm0}\digitwidth=\wd0
\newdimen\minuswidth\setbox0=\hbox{\mathsurround=0pt$-$}\minuswidth=\wd0
{\catcode`?=\active\catcode`+=\active
 \gdef\spdef{\offinterlineskip
 \catcode`?=\active\def?{\kern\digitwidth}%
 \catcode`+=\active\def+{\mathbin{\hbox to\minuswidth{\hfil}}}}}
	\tableitem{conduct} The electrical conductivity for various
values of the ion charge $Z_i$ and for various electron-electron
collision operators.  The conductivities are normalized to
$n_eq_e^2/m_e\nu_{te}$.
	$$\vcenter{\tabskip.5em\spdef
\halign{\strut\hfil\rm #\hfil\tabskip2em&\hfil$# $\hfil
&\hfil$# $\hfil&\hfil$# $\hfil&\hfil$# $\hfil
\tabskip.5em\cr\noalign{\hrule\vskip1.5pt\hrule}
Collision operator&Z_i=1&Z_i=2&Z_i=5&Z_i=10\cr
\noalign{\hrule}
linearized        &   7.429  &   4.377  &   2.078  &   1.133  \cr
drifting          &   6.331  &   3.876  &   1.932  &   1.084  \cr
Maxwellian        &   3.773  &   2.824  &   1.660  &   0.998  \cr
high-velocity     &   2.837  &   2.310  &   1.489  &   0.938  \cr
\noalign{\hrule\vskip1.5pt\hrule}}}$$
	\tableitem{rel-conduct} The electrical conductivity of a
relativistic plasma for various values of the ion charge $Z_i$ and for
various electron temperatures.  The conductivities are normalized to
$n_eq_e^2/m_e\nu_{te}$ and the electron temperatures are given in terms
of $\Theta=T_e/m_ec^2$.
	$$\vcenter{\tabskip.5em\spdef
\halign{\strut\hfil$# $\hfil\tabskip2em&\hfil$# $\hfil
&\hfil$# $\hfil&\hfil$# $\hfil&\hfil$# $\hfil
\tabskip.5em\cr\noalign{\hrule\vskip1.5pt\hrule}
\Theta&Z_i=1&Z_i=2&Z_i=5&Z_i=10\cr
\noalign{\hrule}
0.0?      &   7.429  &   4.377  &   2.078  &   1.133  \cr
0.01      &   7.291  &   4.275  &   2.019  &   1.097  \cr
0.02      &   7.160  &   4.180  &   1.963  &   1.064  \cr
0.05      &   6.807  &   3.928  &   1.821  &   0.979  \cr
0.1?      &   6.317  &   3.590  &   1.636  &   0.872  \cr
0.2?      &   5.575  &   3.102  &   1.383  &   0.729  \cr
\noalign{\hrule\vskip1.5pt\hrule}}}$$
	\tableitem{fortran-vars}  Meaning of {\sc FORTRAN}
variables and arrays.
	$$\vcenter{\tabskip.5em
\halign{\strut$# $\hfil\tabskip2em&$# $\hfil
\tabskip.5em\cr\noalign{\hrule\vskip1.5pt\hrule}
\hbox{\rm{\sc FORTRAN} name}&\hbox{\rm meaning}\cr
\noalign{\hrule}
dx&\Delta v\cr
dy&\Delta\theta\cr
dt&\Delta t\cr
jx&N\cr
iy&M\cr
xg(j)&v_j\cr
x(i)&v_{j+1/2}\cr
yg(i)&\theta_i\cr
y(i)&\theta_{i+1/2}\cr
cg(i)&\cos\theta_i\cr
cn(i)&\cos\theta_{i+1/2}\cr
sg(i)&\sin\theta_i\cr
sn(i)&\sin\theta_{i+1/2}\cr
pi&\pi\cr
f(i,j)&f_{i+1/2,j+1/2}\cr
{\it fm}(j)&f_{m,j+1/2}\cr
cur&\numint (v\cos\theta)\cr
{\it f1}(j)&f^{(1)}(v_{j+1/2})\cr
{\it c1}(j)*cn(i)&\left.C\bigl(f_m(v),f^{(1)}(v)\cos\theta\bigr)\right|_{i+1/2,j+1/2}\cr
ax(i,j)&a_{v,i+1/2,j+1/2}\cr
ay(i,j)&a_{\theta,i+1/2,j+1/2}\cr
phi(i,j)&\phi_{i+1/2,j+1/2}\cr
\noalign{\hrule\vskip1.5pt\hrule}}}$$
\end{thetables}

\begin{thefigures}{99}

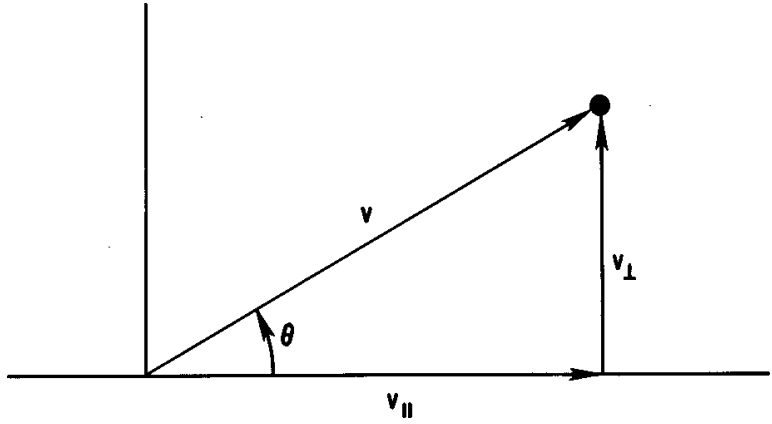
\figitem{coord-fig}{4in} The cylindrical and spherical coordinate systems.

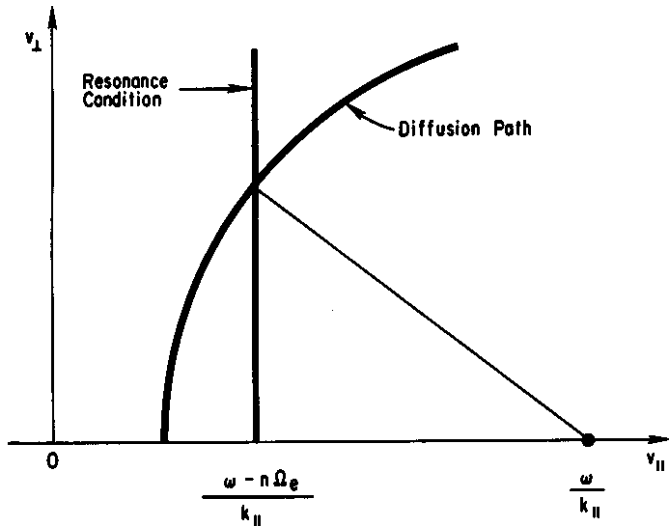
\figitem{diff-fig}{3.5in} The relation between the resonance condition for
quasilinear diffusion $\omega-k_\parallel v_\parallel-n\Omega_e=0$ and
the diffusion path $({\bf v}-(\omega/k_\parallel){\bf\hat
v}_\parallel)^2={\rm const}$.

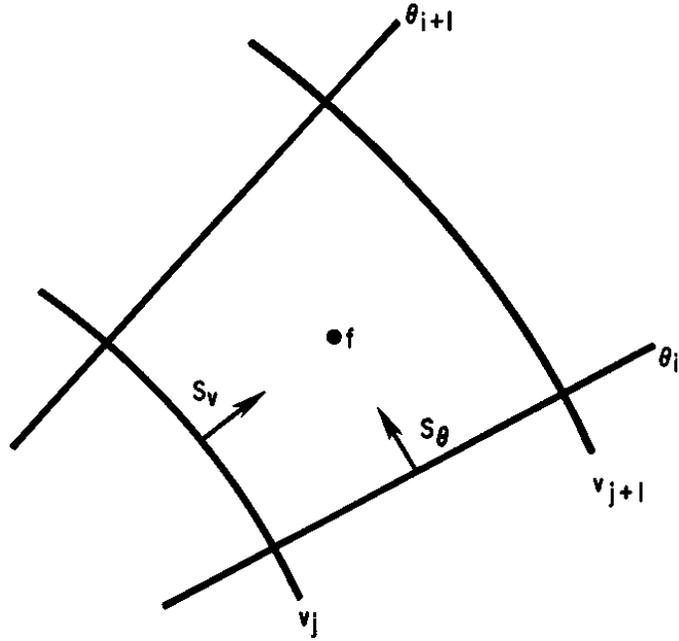
\figitem{grid-fig}{3.5in} The numerical grid showing where the distribution
function and the fluxes are defined.

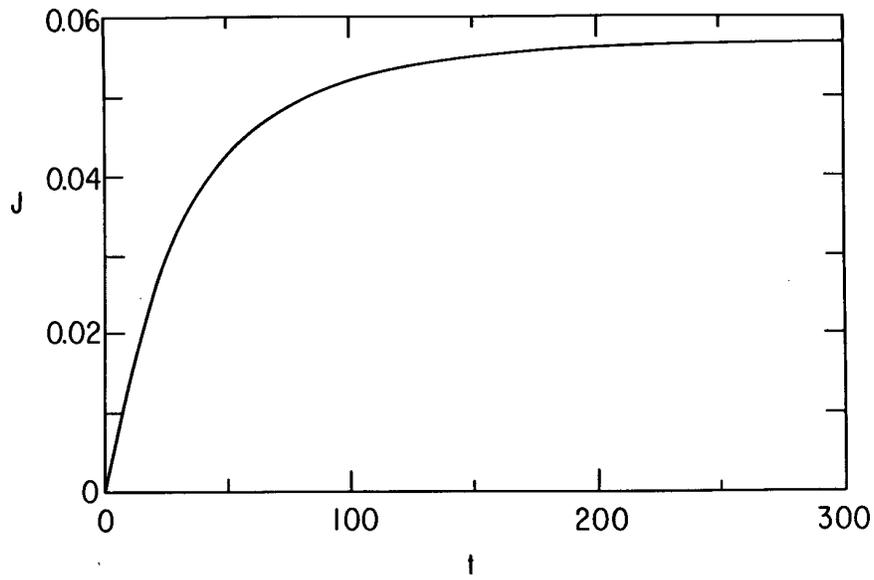
\figitem{current}{4.5in} The current as a function of time for $Z_i=1$,
$f(t=0)=f_m$, and rf diffusion given by \eqsref{d-ql} with $D_0=1$,
$v_1=3$, and $v_2=5$.  Here we have $M=N=100$, $\Delta t=0.2$, and
$v_{\rm max}=10$.  Electron-electron collisions are computed using
$C_{\rm Max}^{e/e}$.

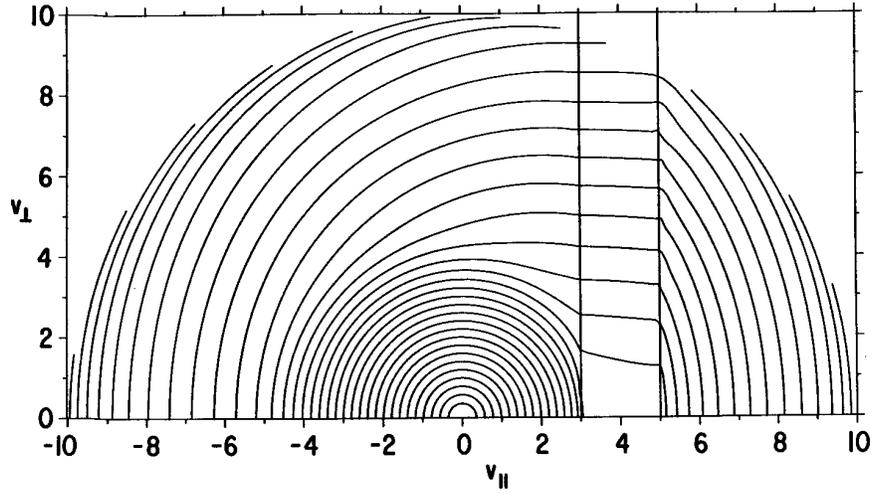
\figitem{f-steady}{4.5in} The steady-state distribution for the case shown in
\figref{current}.  The contour levels are $f=(2\pi)^{-3/2}\times
\exp[-\half(j/5)^2]$ for $j={\rm integer}$.  This gives equally spaced
contours for a Maxwellian distribution with spacing $\delta v=\fract1/5$.
The resonant region is shown.

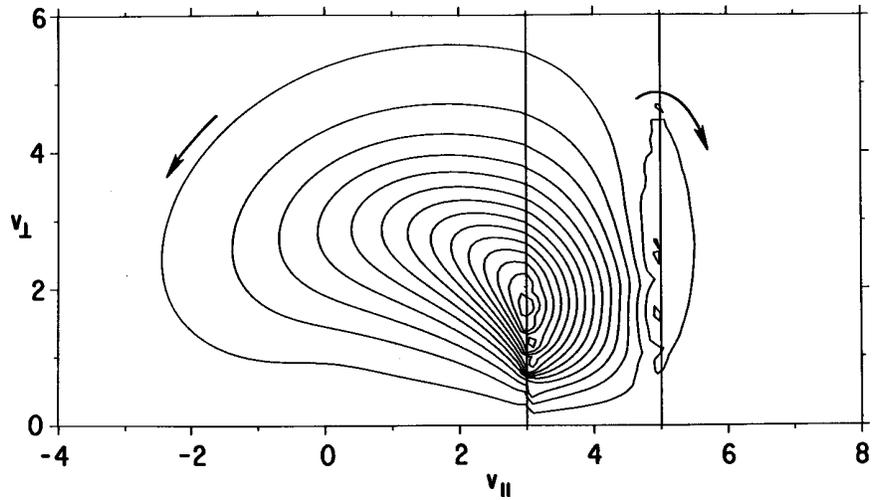
\figitem{flux-fig}{4.5in} The flux plot for the case shown in
\figref{f-steady}.  The plot was obtained by plotting contours of the
stream function $A$, \eqref{stream-grid}.  The contour levels are
$2\times10^{-5}(j+\half)$ for $j={\rm integer}$.

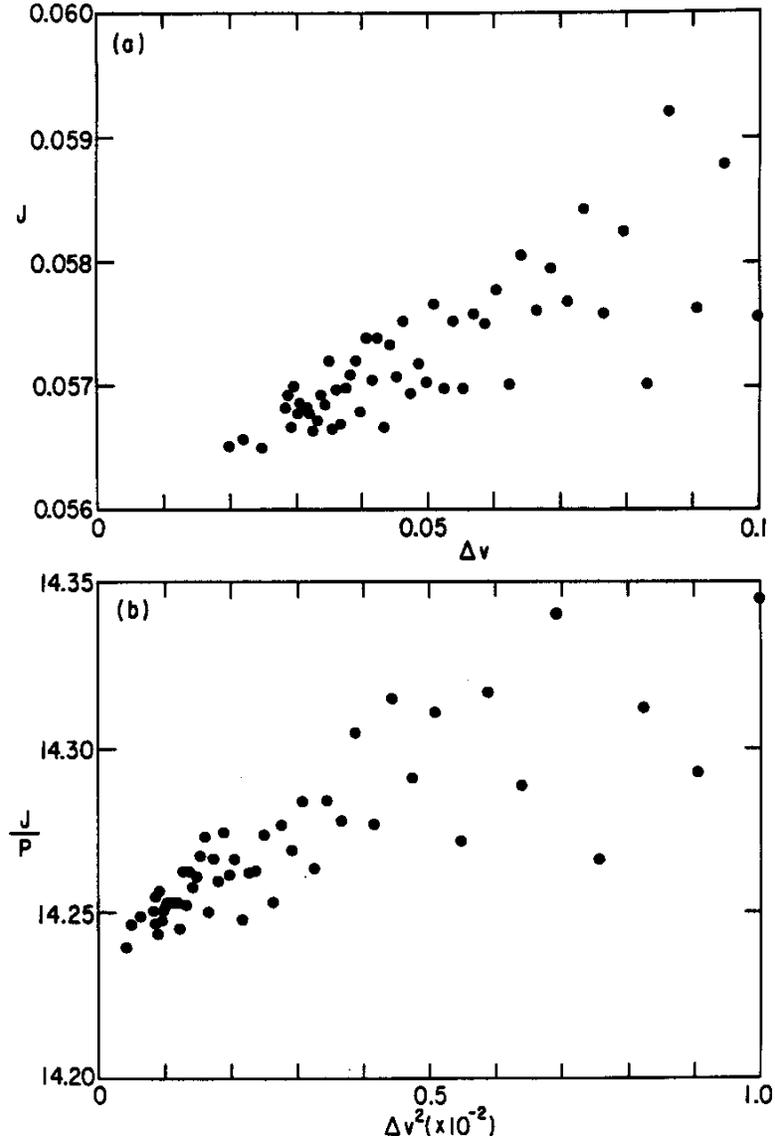
\figitem{conv-d}{4in} The current $J$ (a) and the efficiency $J/P$ (b) as
functions of $\Delta v$.  The parameters are the same as for
\figref{current} except that $M$ and $N$ are allowed to vary with
$M=N$.  The plots show the results from runs with $N$ varying between
100 and 350 in steps of 5 and between 350 and 500 in steps of 50.

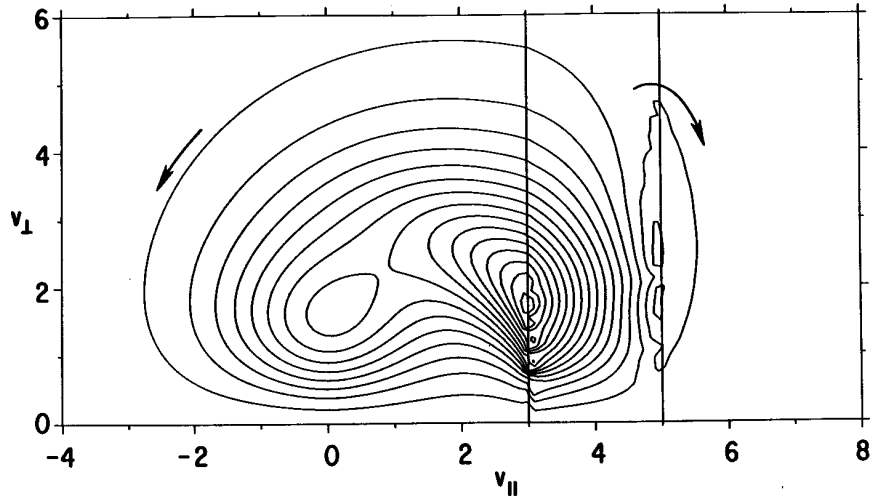
\figitem{trunc-flux}{4.5in} The flux plot when $C_{\rm trunc}^{e/e}$ is used.
The parameters are otherwise the same as for \figref{flux-fig}.

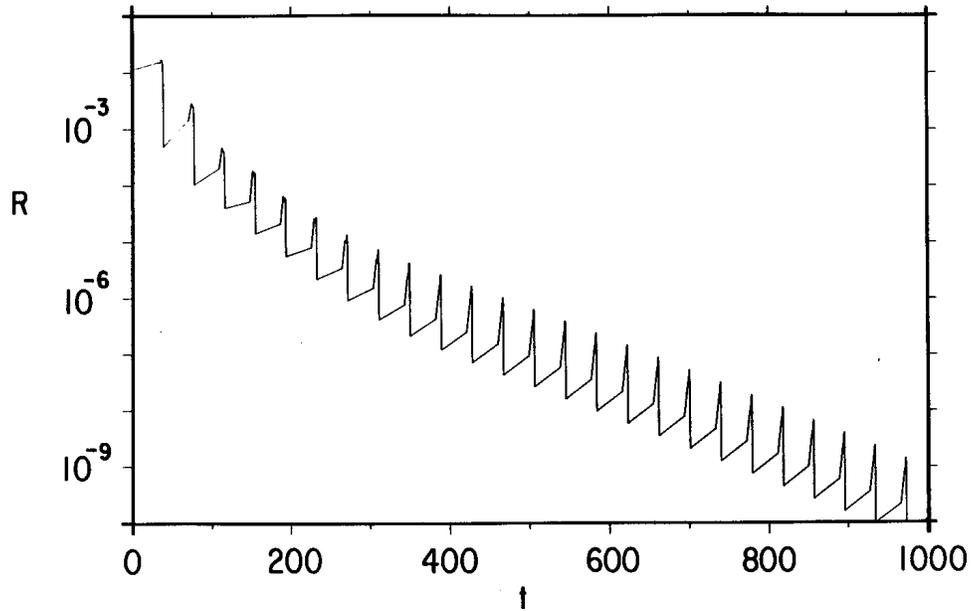
\figitem{residu-fig}{5in} $R$ as a function of time when Chebyshev
acceleration is applied to the example shown in \figref{f-steady}.  Here
$1/\beta=0.05$, $1/\alpha=1000$, $K=20$, and $f(t=0)=f_m$.  The
convergence criterion $R<10^{-9}$ is met after 400 steps at $t=790$.

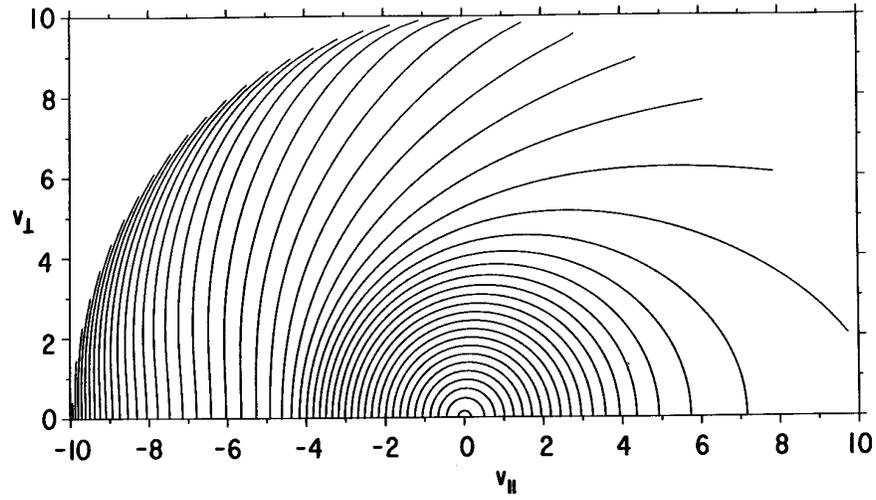
\figitem{run-fig}{4.5in} The steady-state distribution in the presence of a dc
electric field.  Here we have $Z_i=1$, $E=0.06$, $M=N=100$, $v_{\rm
max}=10$ and electron-electron collisions are computed using $C_{\rm
Max}^{e/e}$.  The contour levels are the same as for \figref{f-steady}.

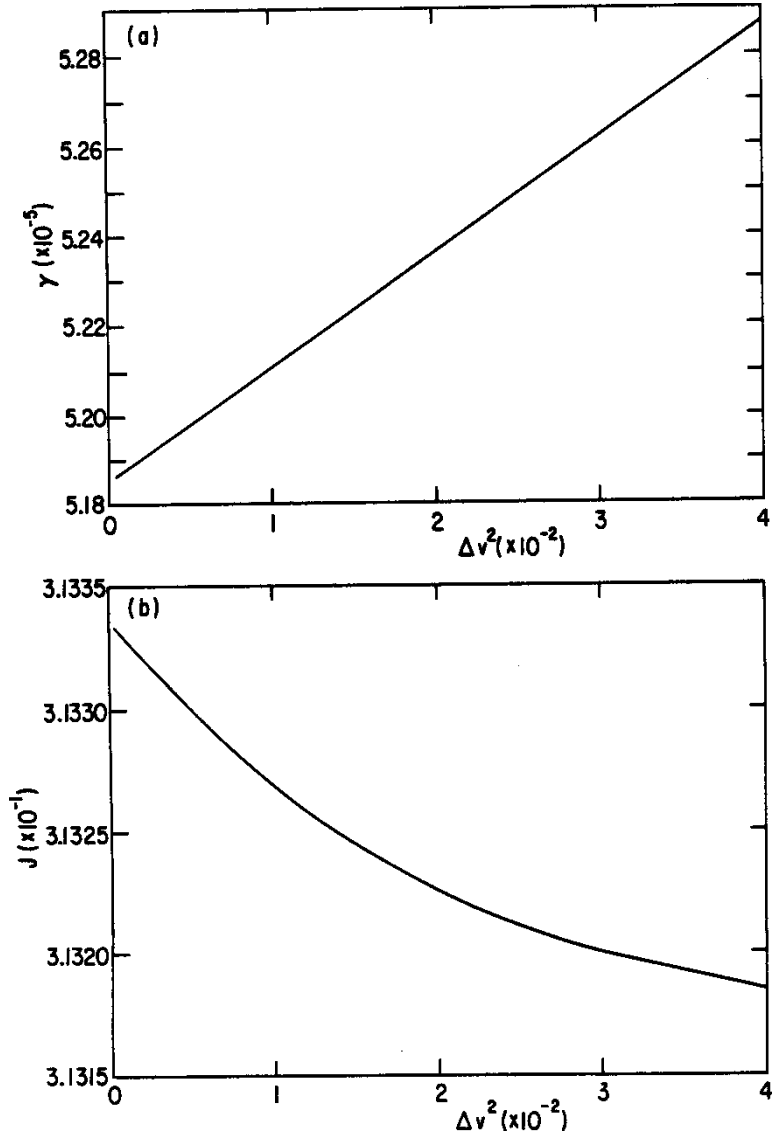
\figitem{conv-e}{4in} The runaway rate $\gamma$ (a) and the current $J$ (b)
as functions of $\Delta v$.  The parameters are the same as for
\figref{run-fig} except that $M$ and $N$ are allowed to vary with
$M=N$.  The plots show the results from runs with $N$ varying between
50 and 300 in steps of 10 and between 300 and 500 in steps of 50.

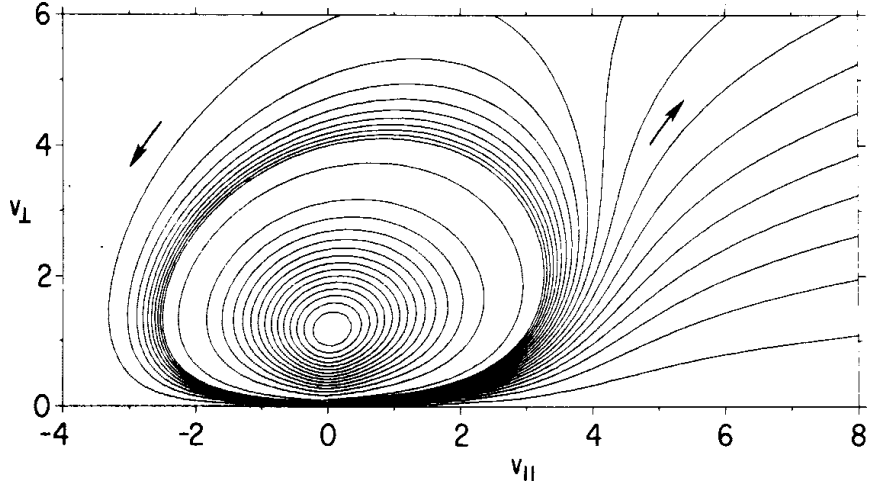
\figitem{run-flux}{4.5in} The flux plot for the runaway problem.  This
illustrates the same case as shown in \figref{run-fig} except that a
source of particles is introduced at the origin to balance the runaway
loss $\gamma=5.148\times10^{-5}$.  One set of contour levels is
$0.1\gamma(j+\half)$ for $j={\rm integer}$ and $-10\le j <10$ (these
give the stream lines that run away and the outermost stream lines that
encircle the central eddy).  The other set of contour levels is
$2\times10^{-4}(j+\half)$ for $j={\rm integer}$ and $j>0$ (these are
the innermost stream lines about the eddy).

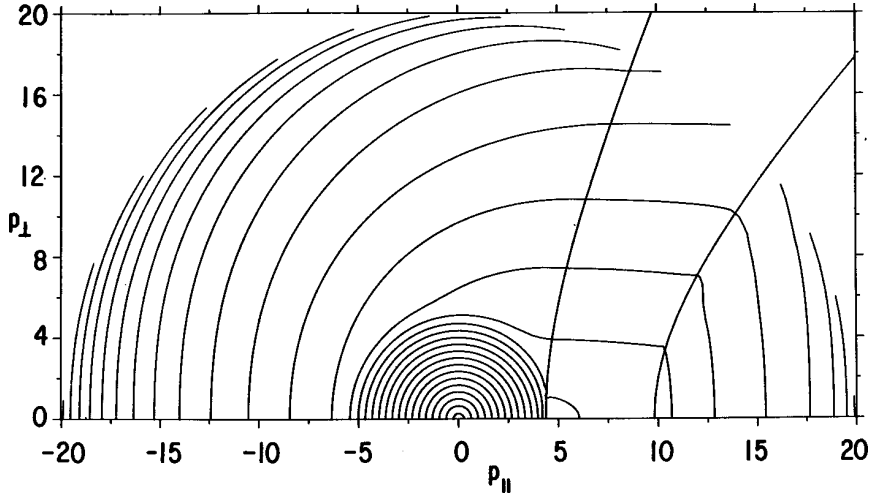
\figitem{rel-fig}{4.5in} The steady-state distribution for $Z_i=1$,
$\Theta=0.01$ ($T_e=5.11\,\rm keV$), and rf diffusion given by
\eqref{d-ql-rel} with $D_0=1$, $v_1=0.4c$, and $v_2=0.7c$.  Here we have
$M=N=100$ and $p_{\rm max}=20$.  Electron-electron collisions are
computed using $C_{\rm trunc}^{e/e}$.  The contour levels are chosen to
be $f=\sqrt\Theta \exp[-\sqrt{1+\Theta(j/3)^2}/\Theta]/[4\pi
K_2(\Theta^{-1})]$ for $j=\rm integer$ which give equally spaced
contours for a relativistic Maxwellian with spacing $\delta
p=\fract1/3$.  [For $\Theta=0.01$ we have
$K_2(\Theta^{-1})=1.019\sqrt{\pi/2}\sqrt\Theta\exp(-1/\Theta)$.]  The
resonant region is shown.

\end{thefigures}
\end{document}